\documentclass[10pt, conference, letterpaper]{ IEEEtran}
\IEEEoverridecommandlockouts

\usepackage{xcolor}
\def\BibTeX{{\rm B\kern-.05em{\sc i\kern-.025em b}\kern-.08em
    T\kern-.1667em\lower.7ex\hbox{E}\kern-.125emX}}

\usepackage{booktabs}
\usepackage{siunitx}
\usepackage{hyperref}

\newtheorem{definition}{Definition}
\newtheorem{protocol}{Protocol}
\usepackage{cite}
\usepackage{amsmath,amssymb,amsfonts}
\usepackage{graphicx}
\usepackage{textcomp}
\usepackage{xcolor}
\usepackage{booktabs}
\usepackage{siunitx}
\usepackage{subcaption}
\usepackage[ruled,vlined,linesnumbered]{algorithm2e}
\usepackage{caption}

\begin{document}

\title{NSMA: Neuro-Symbolic Manifold Alignment for Generalizable Adaptive Bitrate Streaming under Texture Shift}

\author{\IEEEauthorblockN{Zhiqiang He}
\IEEEauthorblockA{\textit{The University of Electro-Communications}\\
Tokyo, Japan \\
hezhiqiang@ieee.org}
\and
\IEEEauthorblockN{Zhi Liu}
\IEEEauthorblockA{
\textit{The University of Electro-Communications}\\
Tokyo, Japan \\
liu@ieee.org}
}


\maketitle

\begin{abstract}


For decades, ABR has kept two kinds of intelligence apart. Neural policies learn rich behaviors yet forget them the moment the environment changes; rules never learn, and never forget. Every prior attempt to combine them has kept this separation, letting rules supervise, constrain, or override the network from outside. We dissolve the boundary itself. But no union can be trusted before it can be tested, and ABR has never known how to measure what its policies learn or forget. The field's yardstick is bandwidth statistics, and we show it misleads. Identical statistics can hide entirely different outcomes, while wildly different statistics can hide similar ones. We replace the yardstick before building the bridge, with Texture-Aware Generalization Evaluation, a protocol that judges a policy by its whole training journey across traces whose temporal nature is laid bare. What truly breaks a policy is invisible. No statistic reveals it, no feature extracts it, yet rules walk through it untouched, for they reason from physics and owe the data nothing. So we build the bridge. Neuro-Symbolic Manifold Alignment (NSMA) embeds rule decisions as anchors inside the latent space of the neural policy, so that it keeps learning where learning pays, and can no longer forget what rules have always known. Generalization cannot be argued, only survived. We raise NSMA on 3G traces alone and release it, without fine-tuning, into eight unseen datasets spanning 4G, 5G, and WiFi, and onto a real-world player. It outperforms every state-of-the-art baseline. And when we open its latent space to ask why, probing and visualization return the same answer the design promised. \url{https://tinyzqh.github.io/NSMA/}




\end{abstract}

\begin{figure*}
  \includegraphics[width=1.35in]{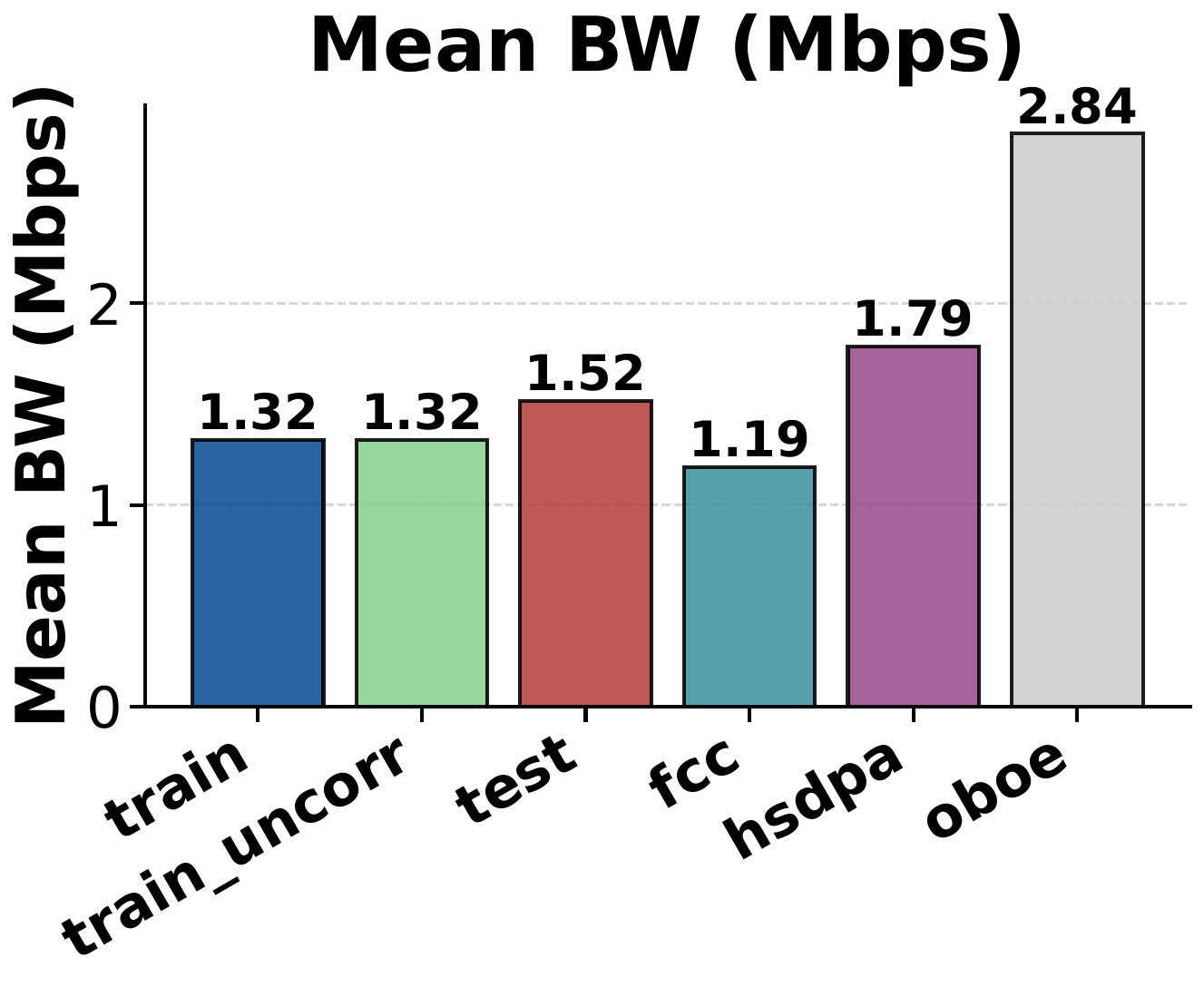}
  \includegraphics[width=1.35in]{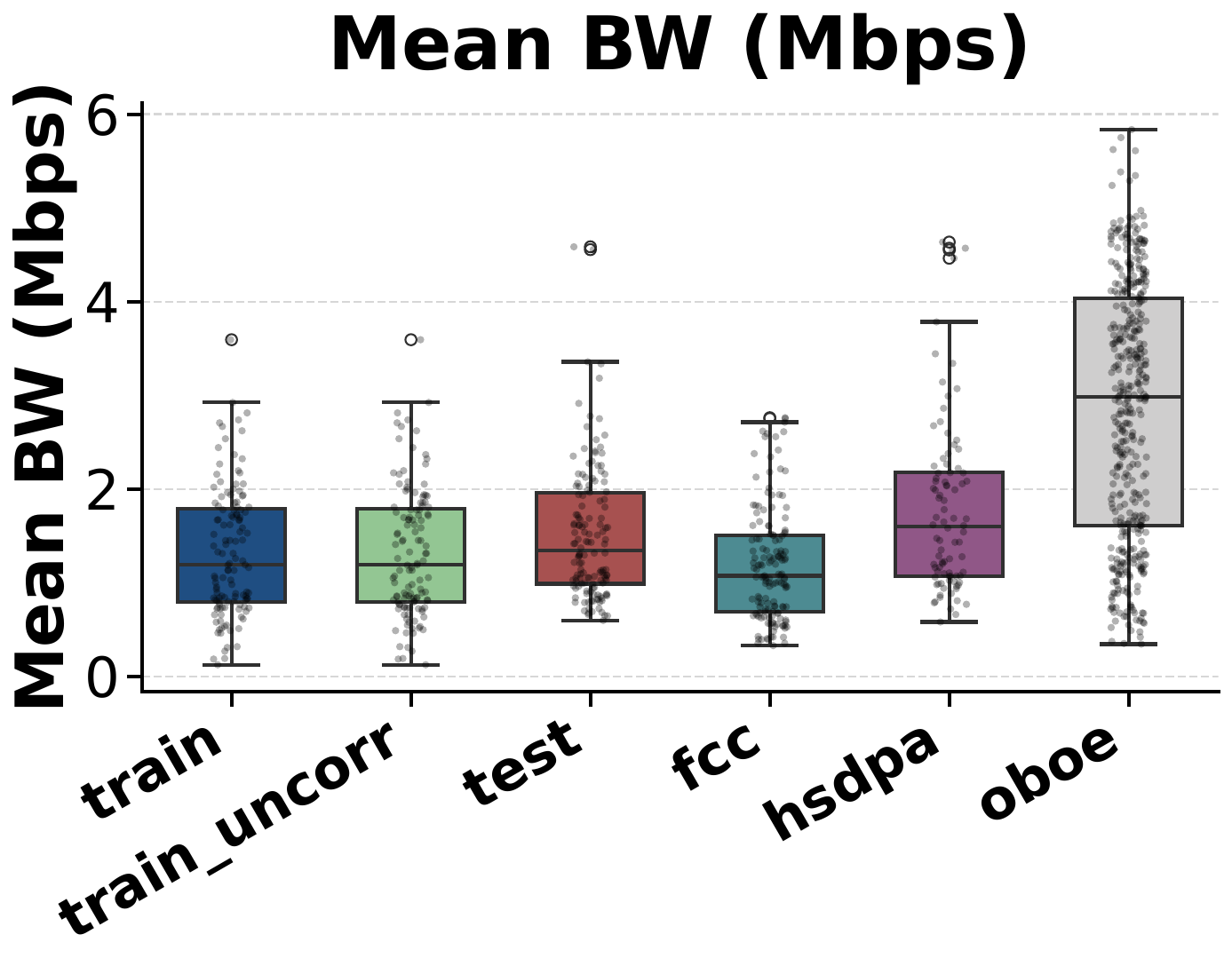}
  \includegraphics[width=1.35in]{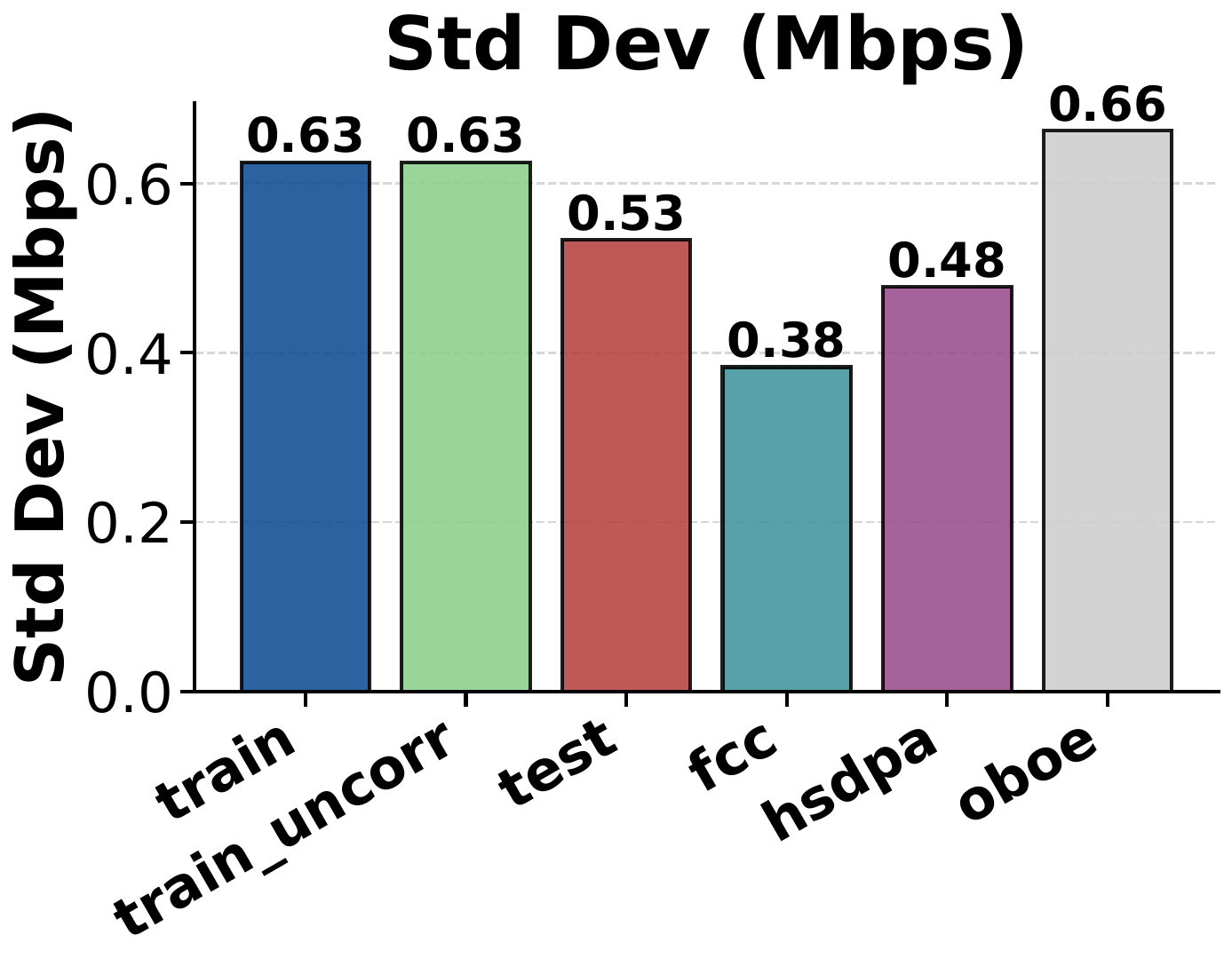}
  \includegraphics[width=1.35in]{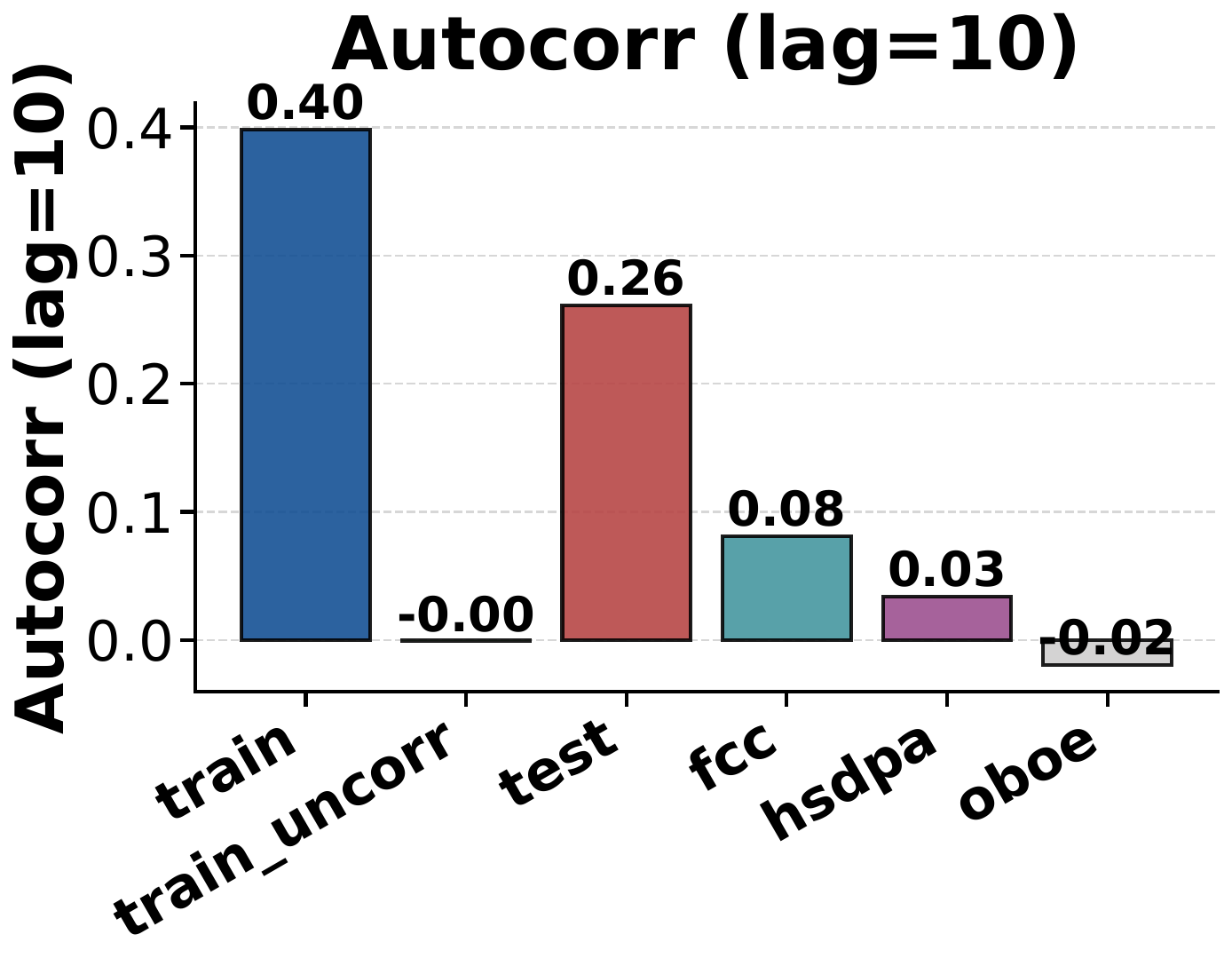}
  \includegraphics[width=1.35in]{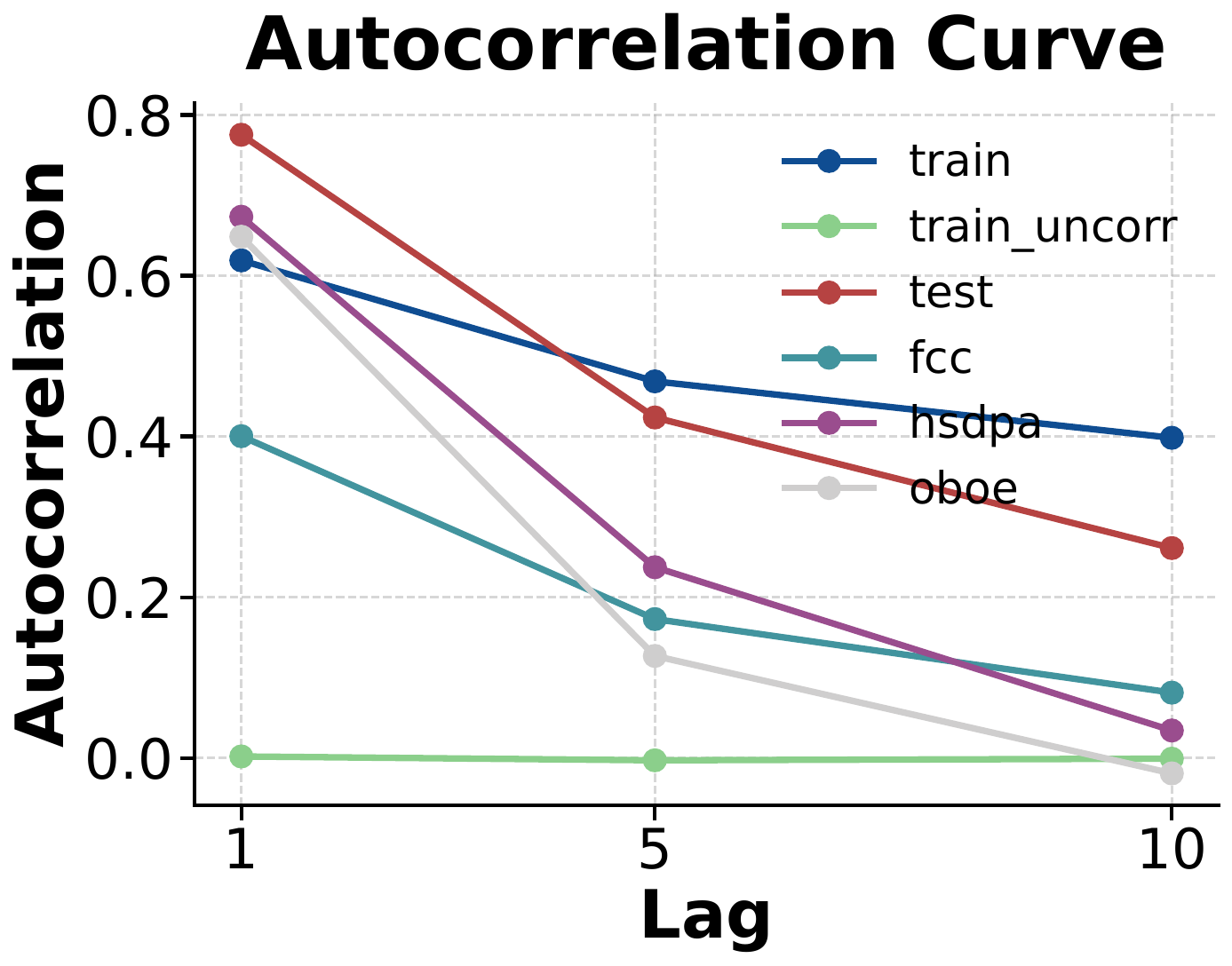}
  \includegraphics[width=1.35in]{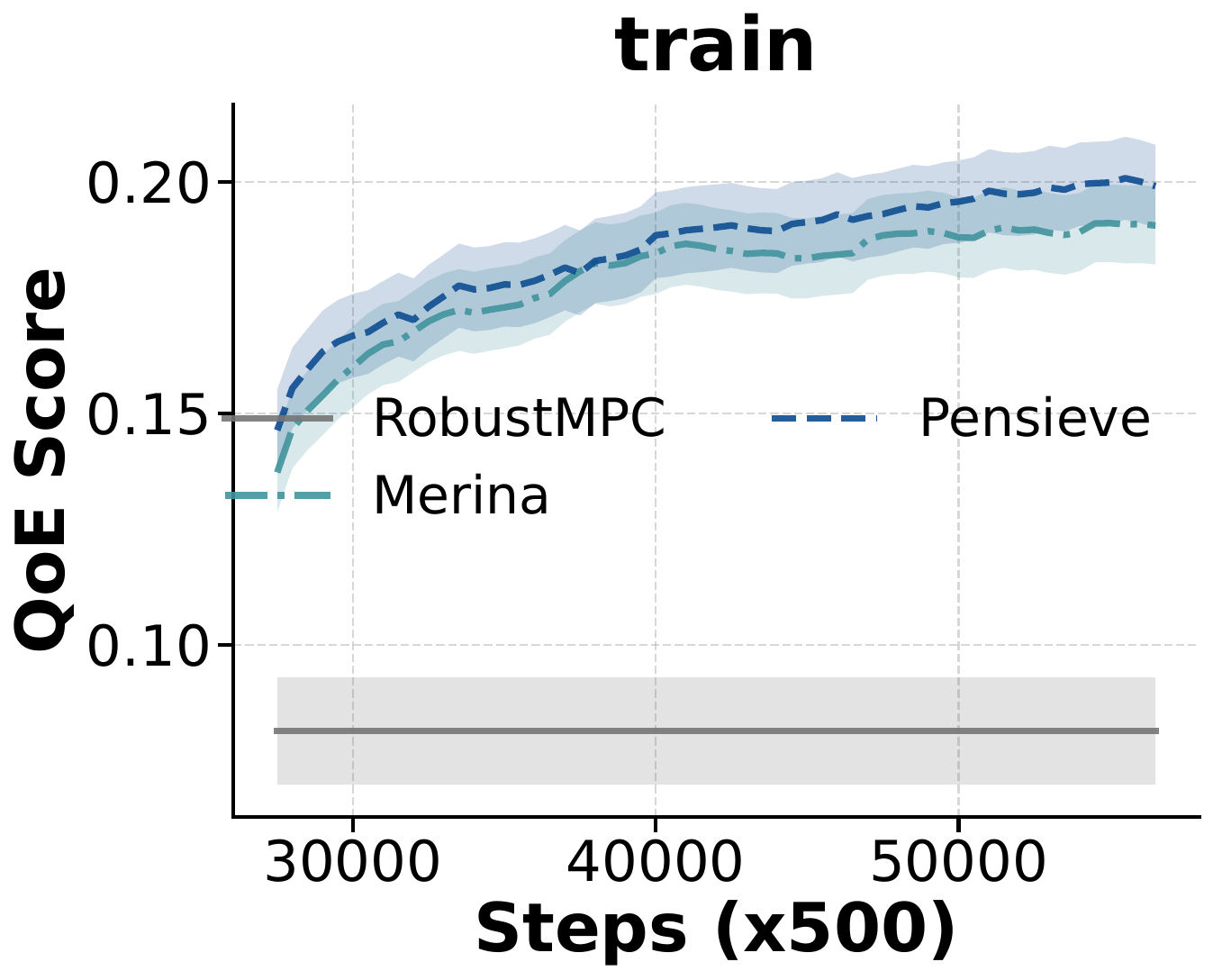}
  \includegraphics[width=1.35in]{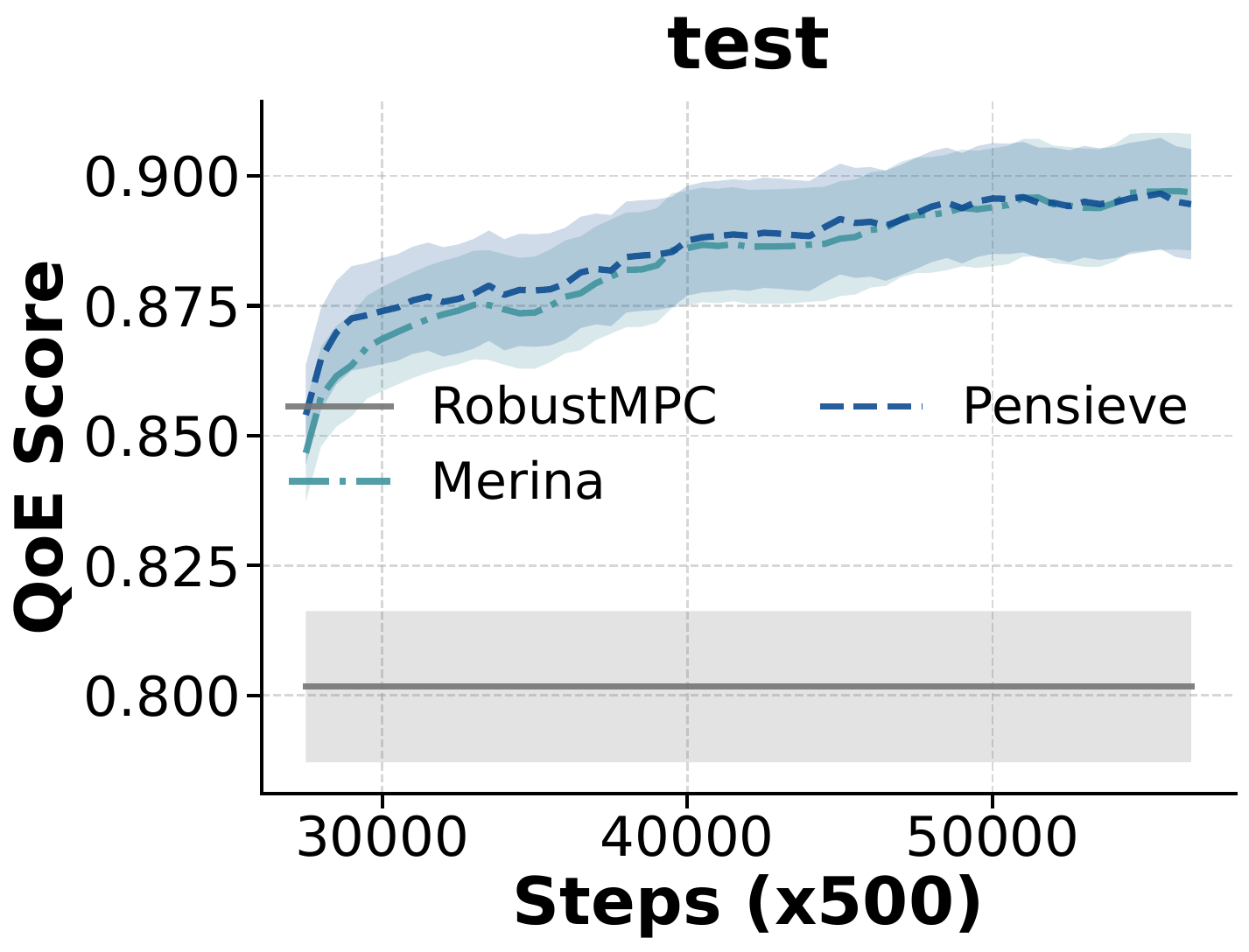}
  \includegraphics[width=1.35in]{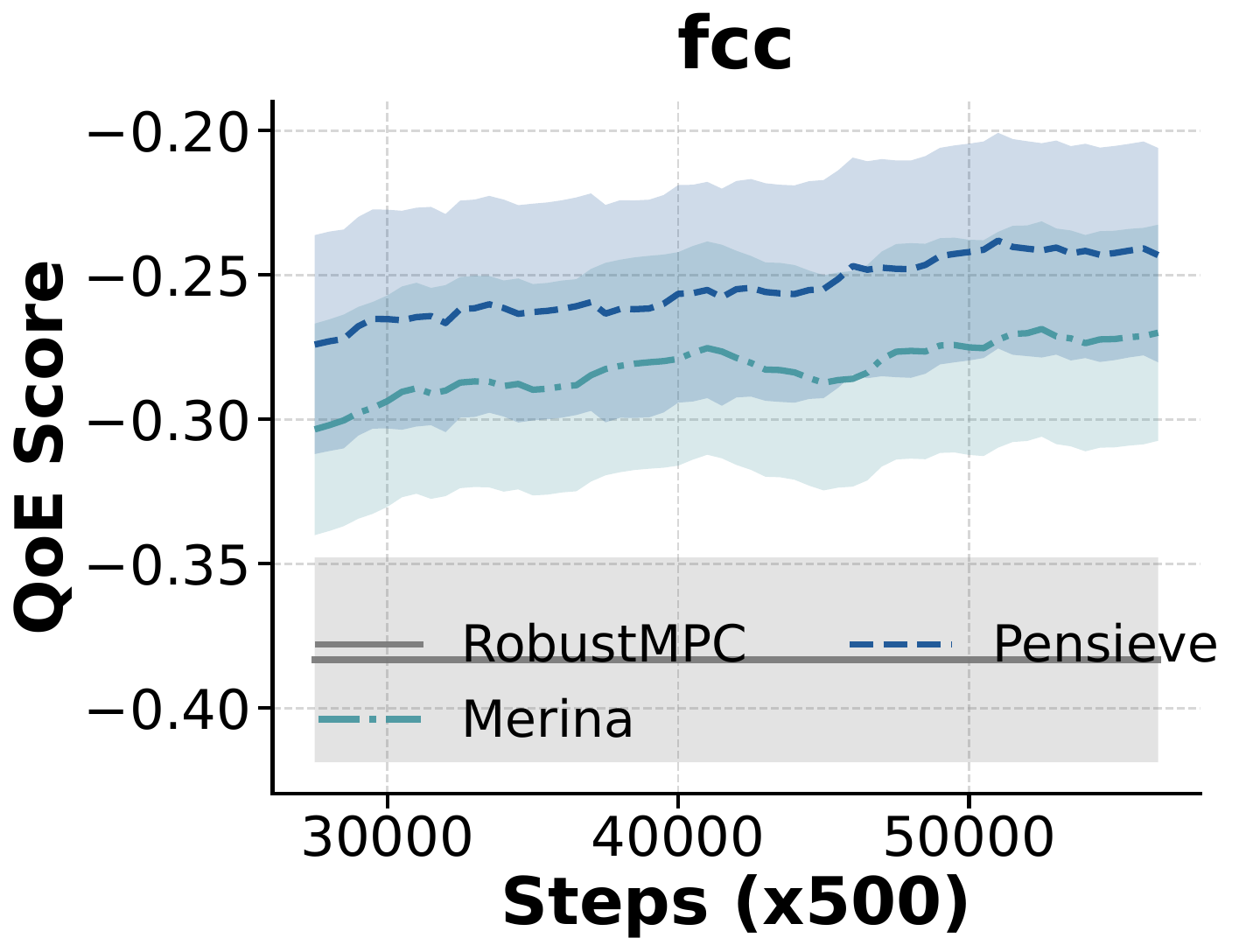}
  \includegraphics[width=1.35in]{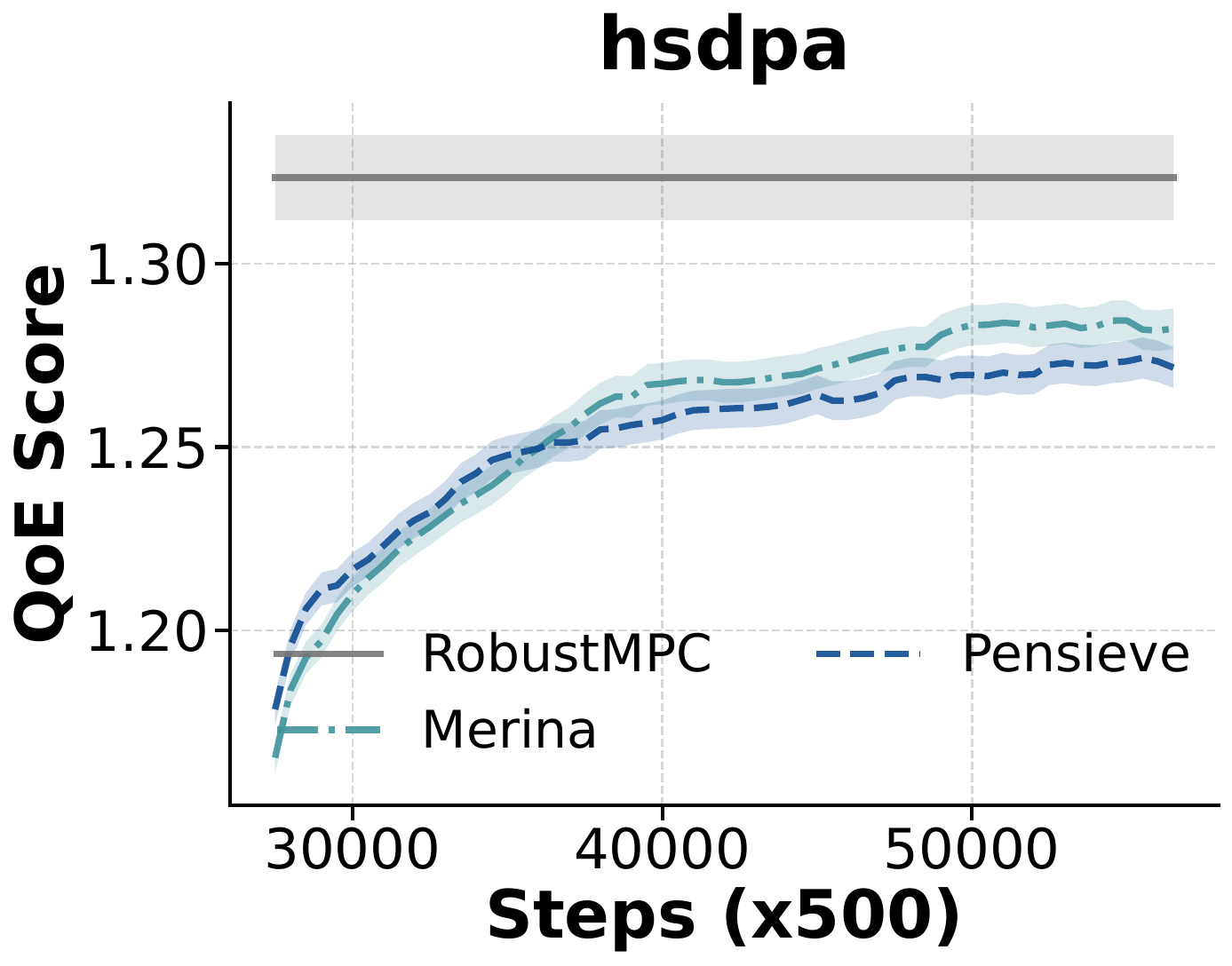}
  \includegraphics[width=1.35in]{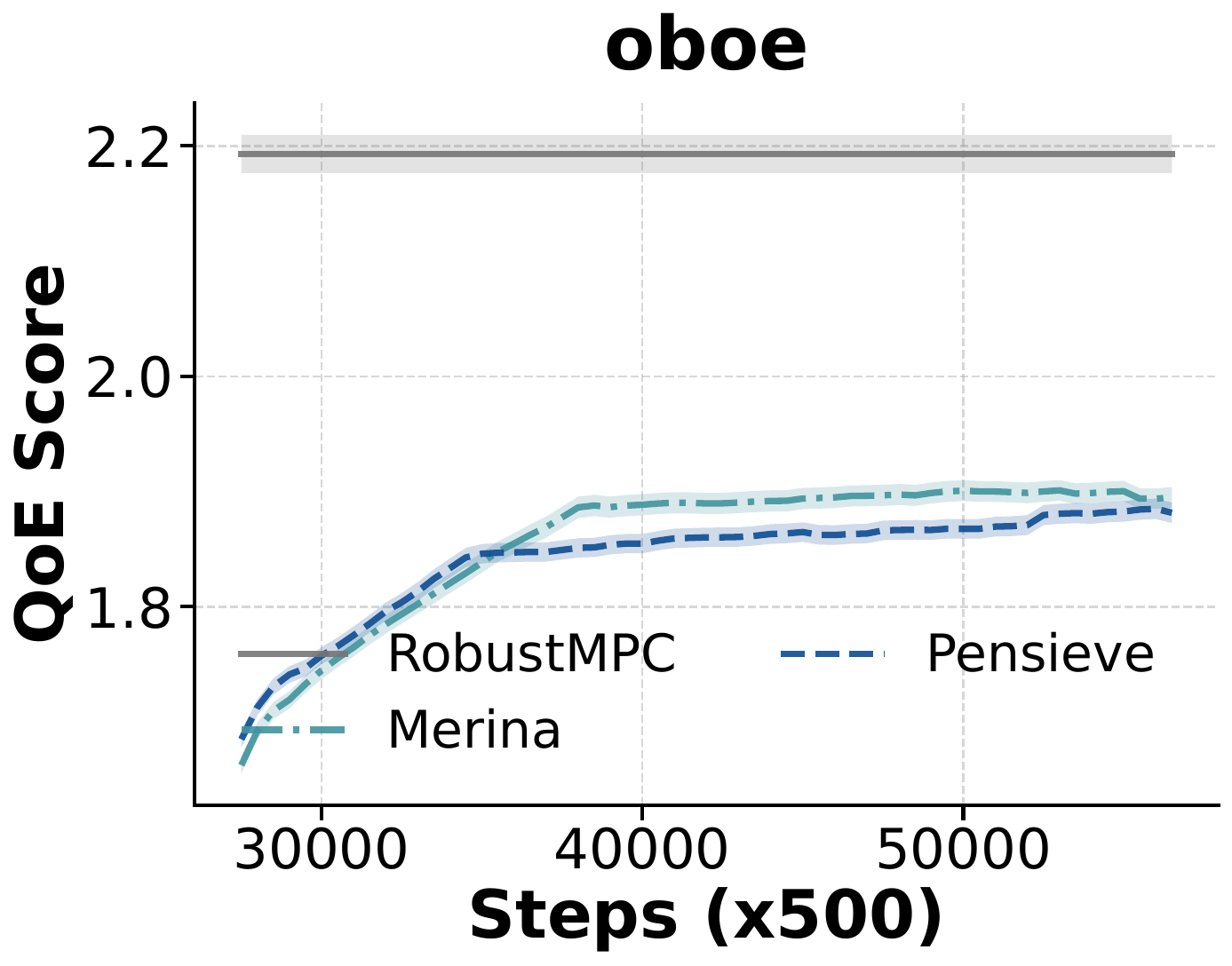}%
  \caption{Dataset identity does not predict generalization. \emph{Top:} statistical profiles of six trace environments: mean bandwidth, distribution, standard deviation, and temporal autocorrelation. \emph{Bottom:} QoE of learning-based agents (Pensieve, MERINA) versus RobustMPC (gray band); Agents are trained solely on \texttt{train}, and every other panel evaluates the saved training weights on an unseen environment. \texttt{fcc} is statistically far from \texttt{train} yet performs like it, the very pattern past work has celebrated as generalization; \texttt{train\_uncorr} raises no statistical alarm yet collapses (Fig.~\ref{train_uncorr}).}
  \label{trace_question}
\end{figure*}

\begin{IEEEkeywords}
Adaptive Video Streaming, Generalization, Reinforcement Learning, Manifold Alignment, Neuro-Symbolic, Evaluation Protocol.
\end{IEEEkeywords}

\section{Introduction}
Across networked systems, there has been a quiet competition for decades between two decision-making methods. Hand-crafted algorithms, distilled from domain principles, are trusted but conservative; learning-based methods, trained from data, are powerful within the worlds they have seen and fragile beyond them. Adaptive bitrate (ABR) streaming compresses this contest into its purest form. As the dominant source of Internet traffic \cite{10752828}, an ABR algorithm must make a bitrate decision every few seconds, without knowing what the network will do next \cite{he2025understanding}. Each commitment poses the question the field has never settled. Trust what the data taught, or hold to what the principles say. This paper answers by refusing the choice, dissolving the boundary between the two.

Efforts to combine the two are not a new concept, and they differ in their placement of the rule, which is quite revealing. Imitation and warm-starting \cite{huang2019comyco} place it before the network as a teacher, whose lessons fade once training ends and the distribution drifts. Switching and safety constraints \cite{kan2025merina+} place it after as a guard, who inspects the outputs but never touches the representation where failure actually forms. Two-stage designs \cite{peng2019hybrid} place it beside as a colleague, though two logics without shared gradients cannot be jointly optimal \cite{he2025understanding}. Teacher, guard, colleague. In every role the rule remains a second party standing next to the network, because two parties is what everyone assumed. We remove the assumption. Our answer places the rule not beside the network but within it, embedded in the latent space as part of the representation itself.

\textbf{The Illusion of Diversity.} Every conclusion in this field, on rules, on learning, on their every combination, comes from the same kind of evaluation: testing across trace datasets that differ in name, origin, and statistics. This practice is so common that no one thinks to question it. We did, and it failed in both directions. \texttt{fcc} differs from our training set in source and in statistics, yet agents keep their full advantage over rule-based baselines on it (Fig.~\ref{trace_question}). \texttt{train\_uncorr} shares every summary statistic with the training set, being merely a reshuffled copy of it, yet on it the advantage disappears (Fig.~\ref{trace_question} and Fig.~\ref{train_uncorr}). The dataset that looks different is harmless; the dataset that looks the same is fatal. So whatever these statistics describe, it is not what makes a trace hard. Something else decides. Names do not record it, statistics do not see it, yet it is real enough to erase a trained policy's entire advantage. We cannot describe it directly, for its dimensions are open-ended. But we can say precisely where it lives: in everything about a trace that remains free after the statistics are fixed. We call the practice of ignoring it the \textbf{Dataset-Identity Bias}, and the thing itself, \emph{Trace Texture}.
\begin{definition}[Trace Texture]
\label{def:tt}
\emph{Trace Texture} (TT) refers to the intrinsic character of a bandwidth trace beyond what its summary statistics describe, including but not limited to its temporal dynamics, such as autocorrelation, burst patterns, and regime-switching behavior. Two traces are \emph{statistically isomorphic} if they share identical summary statistics; they may nonetheless differ in texture; the two properties are independent.
\end{definition}

\textbf{From texture to a stable signal.} Naming the problem does not solve it. Texture is open-ended and emergent; no feature can capture it. So we stop trying to describe it, and look instead for something immune to it. Such a signal has existed all along. A classical rule never reads a trace. It reasons from buffer dynamics, throughput conservatism, and lookahead constraints. These principles are not blind to texture; they are what decades of expert practice distilled after seeing bandwidth in every form. A rule has learned nothing from data, so a texture shift can take nothing from it. Every decision a rule makes is therefore a stable summary of the environment, unaffected by the texture shifts that break learned policies. The remaining question is where to place this signal. If it enters as an input, the network is free to reweigh or ignore it. If it supervises the output, it never touches the internal representation, which is exactly where texture does its damage. So the signal must be placed inside the representation itself. We put it there.

\textbf{Our approach.} The placement has a name: \textbf{Neuro-Symbolic Manifold Alignment (NSMA)}. Inside NSMA, a \textbf{hyper-connection routing (HCR)} mechanism \cite{zhu2025hyperconnections} encodes the rule's decision and weaves it into the latent space of the neural policy. There it acts as an anchor. The representation around it remains free to learn, but can no longer drift into regions that contradict what the rule knows. The network learns; the anchor holds. Both are trained together and stay differentiable end to end, which is precisely what every earlier hybrid lacked. It is trained on one trace dataset and deployed with no fine-tuning onto eight unseen datasets and a real video player, where it beats state-of-the-art RL and meta-RL baselines outright. And when we open the latent space to check, the anchors are there, holding, exactly as promised.

The core contributions of this work are as follows:
\begin{list}{$\bullet$}{\setlength{\leftmargin}{10.0pt} \setlength{\itemindent}{0.pt} \setlength{\labelsep}{0.5em}}
  \item \textbf{A flaw in the field's measure.} Dataset identity, the names and statistics evaluations rely on, does not predict generalization difficulty. We propose Trace Texture, the character of a trace beyond its statistics.
  \item \textbf{An honest protocol.} Texture-Aware Generalization Evaluation judges a policy over its whole training trajectory, on traces whose texture is reported, not assumed.
  \item \textbf{A unification, not another hybrid.} Prior hybrids place the rule outside the network, as its teacher, guard, or colleague; NSMA places it inside, embedding the rule's decision into the latent space as an anchor of the representation, jointly trained and differentiable end to end.
  \item \textbf{State-of-the-Art Zero-Shot Generalization.} Trained on one dataset, deployed with no fine-tuning, NSMA beats rule-based, RL, and meta-RL baselines on eight unseen environments and a real player, with the anchors verified at work in its latent space.
\end{list}

\section{Background And Motivation}
\subsection{Problem Formulation}

We model ABR streaming as a finite-horizon Markov Decision Process (MDP) over $I$ segment requests, following \cite{yin2015control}. At each step $i$, the agent selects a bitrate $b \in \mathbb{B} = \{b_1, \dots, b_J\}$ for segment $V_i$, which fixes its payload size $d_i^b$ (increasing in $b$). Downloading the segment takes $t_{\mathrm{delay}}(i) = \Big( \tfrac{d_i^b}{S_i} + t_{RTT} \Big)\ \times \eta(i)$, where $S_i$ is the throughput sampled from a bandwidth trace and $\eta(i)$ a multiplicative noise term capturing transmission variability.

Playback drains the buffer while the download proceeds. Let $B_i \in [0, B^{\max}]$ denote the buffer occupancy when segment $V_i$ is requested, and $L$ the playable duration each segment adds. To respect the capacity cap, the client may idle for $\Delta t_i^{w} = \big( (B_i - t_{\mathrm{delay}}(i))_+ + L - B^{\max} \big)_+$ before the next request. The buffer then evolves as $B_{i+1} = \big( (B_i - t_{\mathrm{delay}}(i))_+ + L - \Delta t_i^{w}
\big)_+$, and the next request fires at $t_{i+1}^{d} = t_i^{d} + t_{\mathrm{delay}}(i) + \Delta t_i^{w}$. In short, every bitrate choice is a bet: a larger $d_i^b$ buys quality now at the risk of draining the buffer before the segment lands.

The per-step reward is the increment of QoE, so maximizing return maximizes episode QoE. QoE rewards perceptual quality while penalizing quality oscillation and stalls: $QoE^{I} = \mu_{1} \sum_{i=1}^{I}  q(V_{i}^{b}) - \mu_{2} \sum_{i=1}^{I-1}|q(V_{i + 1}^{b}) - q(V_{i}^{b})| -  \mu_{3} \sum_{i=1}^{I} \left( \left( \frac{d_{i}^{b}}{S_{i}} + t_{RTT} \right) \times \eta(i) - B_{i} \right)_{+}$, where $q(\cdot)$ is a monotone perceptual quality mapping and $\mu_1, \mu_2, \mu_3 \geq 0$ trade off quality, smoothness, and stall avoidance. Following PA-MoE \cite{he2025plasticity} and Merina \cite{kan2025merina+}, we set $\mu_1 = 1$, $\mu_2 = 1$, $\mu_3 = 4.3$.

Following Pensieve, Merina and PA-MoE \cite{mao2017neural, he2025plasticity, kan2025merina+}, the observation $s_i$ stacks six feature channels over the last eight steps: the previous bitrate, buffer occupancy, measured throughput, download delay, the candidate sizes of the next segment, and the number of segments remaining. The action $a_i \in \{0, \dots, 5\}$ indexes the $J = 6$ available encodings. Given $(s_i, a_i)$, the environment samples $(S_i, \eta(i))$, applies the buffer transition, shifts the observation window, and returns the reward.

\subsection{Rethinking Generalization Evaluation}
\label{sec:audit}

\begin{figure}[!htbp]
\centering
\includegraphics[width=\columnwidth]{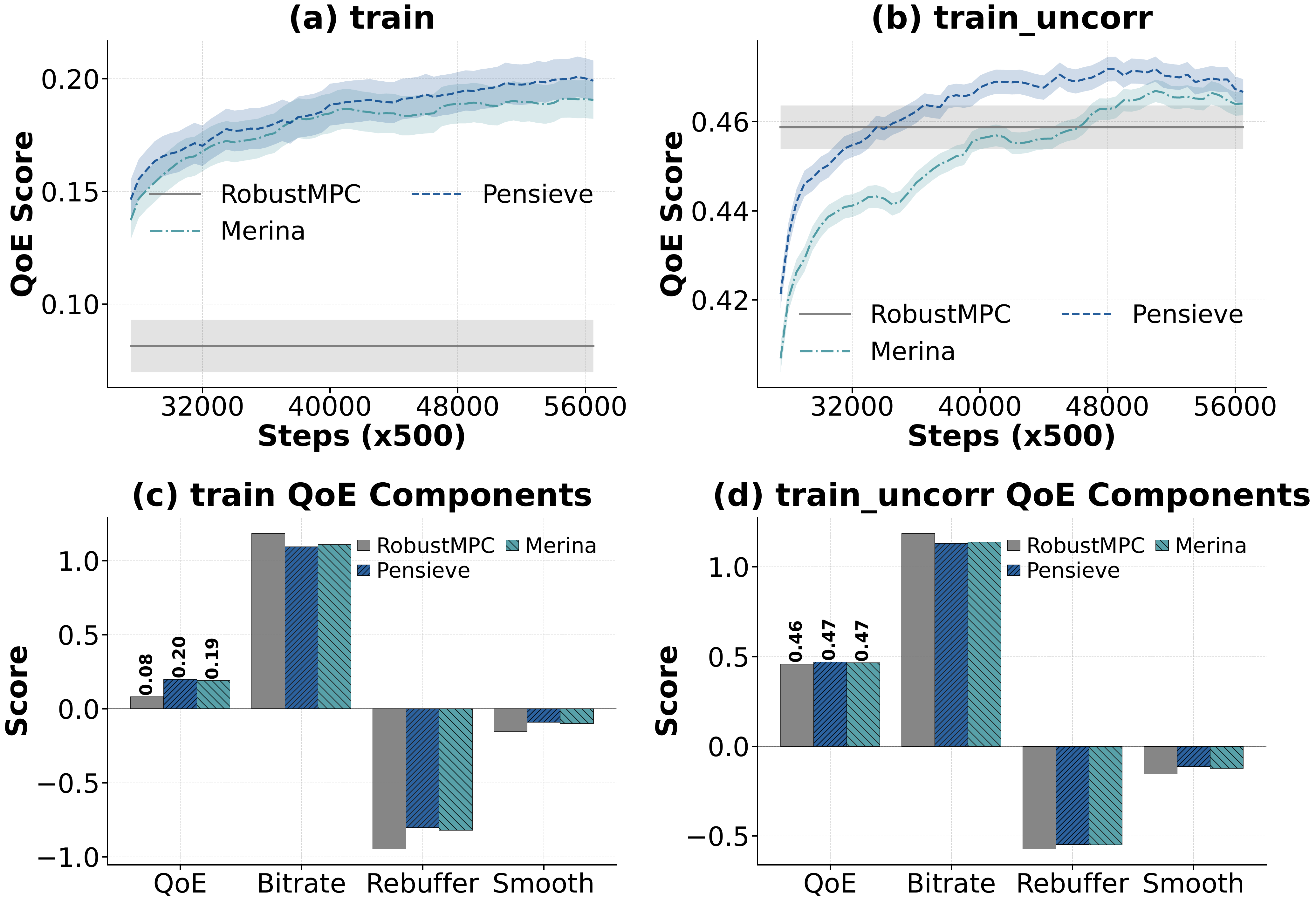}%
\caption{The advantage of learned policies vanishes under texture shift alone.}
\label{train_uncorr}
\end{figure}

Before evaluating any method, ours included, we must settle how generalization should be measured, because the field's current answer does not survive inspection. The standard routine works as follows. A policy is trained on one trace corpus; the checkpoint with the best training QoE is selected; this model is tested on datasets with different names and sources \cite{wu2024mansy, li2023metaabr, kan2025merina+}; and good performance across this ``diversity'' is reported as robustness. The routine relies on two assumptions: (i) that a dataset's identity, its name, origin, and coarse statistics, marks a genuine distribution shift, and (ii) that the checkpoint best on the training set is the right model to examine. Neither has ever been checked. We check both against Fig.~\ref{trace_question} and~\ref{train_uncorr}.

Assumption (i) predicts that \texttt{fcc} should be a demanding test. Its source, mean bandwidth, and variance all differ from the training set (Fig.~\ref{trace_question}, top). The QoE panels say otherwise. Pensieve and Merina hold the same comfortable lead over RobustMPC on \texttt{fcc} that they hold on the training set. The same assumption predicts that \texttt{train\_uncorr} should be trivial. We build it by reshuffling the sequences of the training set, so its mean and variance match the training set exactly. The only thing the reshuffling changes is the temporal structure. Yet on \texttt{train\_uncorr} the agents' large margin over RobustMPC vanishes, converging to the heuristic's level (Fig.~\ref{train_uncorr}). Both predictions fail, and they fail in opposite directions. The dataset judged hard turns out to be easy, and the dataset judged safe breaks the agents. The lesson is not that these two datasets are special. The real problem is what the routine looks at. It checks name, origin, mean, and variance, and none of these tells us how hard a trace actually is.

Assumption (ii) fails for a different reason, and the evidence is already in Fig.~\ref{train_uncorr}. Why is the best-on-train checkpoint the best? Because of its margin over RobustMPC. Fig.~\ref{train_uncorr} shows where that margin comes from. It exists on \texttt{train} and disappears on \texttt{train\_uncorr}, so it is not a general skill but a product of the training texture. The selection is therefore biased from the start. Picking the checkpoint with the largest margin means picking the checkpoint that depends most on the training texture, and then using exactly this model to test texture-independent robustness. The loophole runs deeper than checkpoint selection. Because the routine treats any set of differently named datasets as diverse, one can also select the benchmarks themselves: pick traces that differ in name and statistics but resemble the training set in texture, and the margin is guaranteed to survive. Under the current routine, both the model and the exam can be chosen to flatter each other, and a strong generalization record can be assembled without any generalization taking place.

\begin{protocol}[Texture-Aware Generalization Evaluation]
\label{proto:tage}
A generalization claim for a learning-based ABR policy must meet two requirements: (1) \textbf{Trajectory-based assessment.} The evaluation must be conducted on the sequence of model weights saved throughout training, rather than on a single checkpoint selected for its performance on the training distribution. The results should be reported as performance trajectories for each target environment. (2) \textbf{Trace characterization.} The evaluation must include the marginal statistics and temporal properties identified in Definition~\ref{def:tt} for each environment.
\end{protocol}

A trajectory cannot be cherry-picked. Judged over its entire training run, a policy has nowhere to hide a lucky checkpoint. And the curve tells us more than any single number could, for it shows how generalization is built, whether it holds, and whether it lasts. Trace characterization does for benchmarks what trajectories do for checkpoints. With every environment's temporal character on record, ``diverse'' is no longer a name but a claim, and claims can be checked. The protocol, then, invents nothing. It is the two failures of the old routine, turned into rules. Every evaluation in this paper, our own method's included, follows Protocol~\ref{proto:tage}.

\section{Method}
\label{sec:method}

\begin{figure*}[!htbp]
\centering
\includegraphics[width=\textwidth]{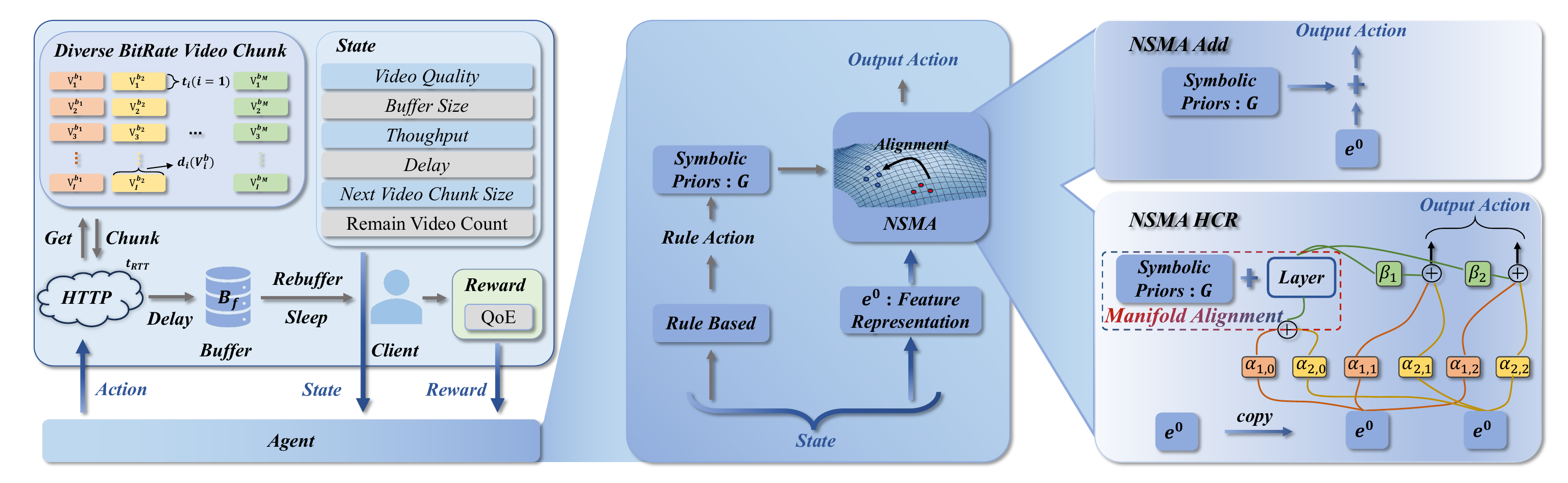}%
\caption{Schematic diagram of the NSMA system. The left panel shows the video streaming state-action-reward cycle, the center illustrates the NSMA agent's internal logic flow, and the right panel compares two structural designs (NSMA Add and NSMA HCR) for embedding logical constraints into the learning process.}
\label{NSMA}
\end{figure*}

The protocol in Section~\ref{sec:audit} tells us whether a policy generalizes. It does not make a policy generalize. To make one, we face the obstacle in Definition~\ref{def:tt}: texture has no fixed list of features, so we cannot extract it, enumerate it, or train against it directly. The way out is to stop chasing texture and use a signal that never depended on it. Rule-based algorithms give us this signal. Their decisions come from transmission physics, not from data, so a texture shift changes nothing for them. The question left is where to put this signal in a neural policy. Feeding it in as an input does not work because the network can learn to ignore it. Supervising the output does not work either, because the damage of texture happens inside the representation, and the output never reaches there to fix it. So we put the signal inside the representation itself. This section builds the mechanism that makes this work.

Our policy follows the standard PPO framework \cite{kan2025merina+, he2025plasticity, wu2024mansy, zhang2025novel}. A neural encoder $\mathbf{P}_e$, adopted from prior work \cite{kan2025merina+, he2025plasticity}, maps the state $s_t$ to a feature embedding $\mathbf{e}^0 \in \mathbb{R}^{d}$ with $d=128$. In parallel, a rule-based agent observes the same state and emits its decision $a_{\mathrm{rule},t} = \mathrm{Heuristic}(s_t)$, which a trainable embedding layer lifts into the same space:
\begin{equation}
\mathbf{G} = \mathrm{Embed}(a_{\mathrm{rule},t}) \in \mathbb{R}^{d}.
\end{equation}
$\mathbf{G}$ is the logic embedding, the texture-immune signal in vector form. Everything that follows is about how $\mathbf{e}^0$ and $\mathbf{G}$ should meet.

\subsection{NSMA-Add: The Simplest Possible Meeting}

The most direct placement is addition:
\begin{equation}
\mathbf{h}_{\mathrm{final}} = \mathbf{e}^{0} + \mathbf{G}.
\end{equation}
We call this variant NSMA-Add. Even this simplest form already does what no prior hybrid did. The rule's signal enters the representation, not the input or the output. And because $\mathrm{Embed}$ is a trainable layer, gradients flow through it, so the rule side and the neural side are trained together, end to end. But addition has an obvious limit. It adds the same vector in the same way, no matter what state the policy is in or how far training has progressed. The policy has no way to use more of the logic in one situation and less in another. NSMA-Add is therefore best understood as the minimal witness that in-representation placement works at all. How well it can work requires giving the meeting a structure.

\subsection{NSMA-HCR: Calibration with a Structure}

Hyper-connection routing (HCR) \cite{zhu2025hyperconnections} replaces the single fixed addition with a routed, state-dependent one. The design has four steps, and each answers a specific need.

\textbf{Step 1: widen the representation.} A single vector offers the logic signal only one place to act. We therefore expand $\mathbf{e}^0$ into $n$ parallel streams,
\begin{equation}
\mathbf{H} = (\mathbf{e}^{0}, \dots, \mathbf{e}^{0})^{T}
\in \mathbb{R}^{n \times d},
\end{equation}
so that the calibrated and the uncalibrated views of the state can coexist and be traded off, rather than overwritten. The streams are normalized as $\bar{\mathbf{H}} = \mathrm{RMSNorm}(\mathbf{H})$ for stable routing.

\textbf{Step 2: let the state decide the routing.} Fixed mixing weights would make the calibration as rigid as NSMA-Add. Instead, the routing weights are generated from the state itself:
\begin{equation}
\mathbf{A} = \bar{\mathbf{A}} + \lambda_{\alpha}
\tanh(\bar{\mathbf{H}}\mathbf{W}_{\alpha}), 
\mathbf{B} = \bar{\mathbf{B}} + \lambda_{\beta}
\tanh(\bar{\mathbf{H}}\mathbf{W}_{\beta})^{T}, \nonumber
\end{equation}
where $\mathbf{W}_{\alpha} \in \mathbb{R}^{d \times (n+1)}$ and $\mathbf{W}_{\beta} \in \mathbb{R}^{d \times 1}$ are learned projections. Two safeguards keep the routing honest. The static baseline $\bar{\mathbf{A}}$ is initialized to identity, so that before training the streams simply pass through; and the bounded $\tanh$ scaled by small $\lambda_{\alpha}, \lambda_{\beta}$ limits early deviations, so the policy cannot wander far from the logic anchor before it has learned anything. The matrix $\mathbf{A} \in \mathbb{R}^{n \times (n+1)}$ splits into $\mathbf{A}_m \in \mathbb{R}^{n \times 1}$, which gathers the streams, and $\mathbf{A}_r \in \mathbb{R}^{n \times n}$, which mixes them laterally.

\begin{algorithm}[hbt!]
\linespread{1.0}\selectfont
\setlength{\parskip}{0pt}
\caption{Forward Pass of NSMA-HCR}
\label{alg:hcr}
\SetKwInOut{Input}{Input}
\SetKwInOut{Output}{Return}
\Input{Feature embedding $\mathbf{e}^0 \in \mathbb{R}^{d}$, logic embedding $\mathbf{G} \in \mathbb{R}^{d}$, number of streams $n$}
\tcp{Step 1: Widen}
Expand into streams: $\mathbf{H} = (\mathbf{e}^0, \dots, \mathbf{e}^0)^{T} \in \mathbb{R}^{n \times d}$\;
Normalize: $\bar{\mathbf{H}} = \mathrm{RMSNorm}(\mathbf{H})$\;
\tcp{Step 2: State-dependent routing}
$\mathbf{A} = \bar{\mathbf{A}} + \lambda_{\alpha} \tanh(\bar{\mathbf{H}} \mathbf{W}_{\alpha}) \in \mathbb{R}^{n \times (n+1)}$\;
$\mathbf{B} = \bar{\mathbf{B}} + \lambda_{\beta} \tanh(\bar{\mathbf{H}} \mathbf{W}_{\beta})^{T} \in \mathbb{R}^{1 \times n}$\;
Split $\mathbf{A}$ into $\mathbf{A}_m \in \mathbb{R}^{n \times 1}$ (gather) and $\mathbf{A}_r \in \mathbb{R}^{n \times n}$ (lateral mix)\;
\tcp{Step 3: Calibrate at the meeting point}
Gather branch input: $\mathbf{h}_0 = \mathbf{H}^{T} \mathbf{A}_m \in \mathbb{R}^{d}$\;
Fuse logic before the transformation: $\mathbf{O} = \mathcal{T}(\mathbf{h}_0 + \mathbf{G}) \in \mathbb{R}^{d}$\;
\tcp{Step 4: Return and settle}
Route back with residual: $\hat{\mathbf{H}} = \mathbf{B}^{T} \mathbf{O}^{T} + \mathbf{A}_r^{T} \mathbf{H} \in \mathbb{R}^{n \times d}$\;
Pool: $\mathbf{h}_{\mathrm{final}} = \sum_{i=1}^{n} \hat{\mathbf{H}}_{i,:} \in \mathbb{R}^{d}$\;
\Output{Consolidated state $\mathbf{h}_{\mathrm{final}}$}
\end{algorithm}

\begin{algorithm}[hbt!]
\caption{PPO with NSMA-HCR}
\label{alg:ppo}
\SetKwInOut{Input}{Input}
\SetKwInOut{Output}{Result}
\Input{Initial policy parameters $\theta_0$, value parameters $\phi_0$, rule-based agent $\mathrm{Heuristic}(\cdot)$, number of streams $n$}
\For{$k = 0, 1, 2, \dots$}{
    \tcp{Stage 1: Rollout with in-representation calibration}
    Initialize empty buffer $\mathcal{D}_k$\;
    \For{each environment step $t$}{
        Encode state: $\mathbf{e}^0_t = \mathbf{P}_e(s_t)$;
        query rule: $\mathbf{G}_t = \mathrm{Embed}(\mathrm{Heuristic}(s_t))$\;
        Consolidate via Algorithm~\ref{alg:hcr}:
        $\mathbf{h}_{\mathrm{final},t} = \mathrm{HCR}(\mathbf{e}^0_t, \mathbf{G}_t, n)$\;
        Sample action $a_t \sim \pi_{\theta_k}(\cdot \mid \mathbf{h}_{\mathrm{final},t})$,
        observe $r_t, s_{t+1}$\;
        Store $(s_t, a_t, r_t, \mathbf{h}_{\mathrm{final},t})$ in $\mathcal{D}_k$\;
    }
    \tcp{Stage 2: Advantage estimation}
    Compute rewards-to-go $\hat{R}_t$ and advantages $\hat{A}_t$ from
    $V_{\phi_k}(\mathbf{h}_{\mathrm{final},t})$\;

    \tcp{Stage 3: Policy and value update}
    \For{$epoch = 1, \dots, K$}{
        Let $r_t^{\mathrm{clip}} = \mathrm{clip}(r_t(\theta), 1{-}\epsilon, 1{+}\epsilon)$\;
        $\theta \leftarrow \arg\max_\theta \, \mathbb{E}_t \big[ \min\big( r_t(\theta)\hat{A}_t,\; r_t^{\mathrm{clip}}\hat{A}_t \big) \big]$\;
        $\phi \leftarrow \arg\min_\phi \,
        \mathbb{E}_t \big[ \big( V_{\phi}(\mathbf{h}_{\mathrm{final},t})
        - \hat{R}_t \big)^2 \big]$\;
    }
}
\Output{Optimized policy $\pi_{\theta}$ and value function $V_{\phi}$}
\end{algorithm}

\textbf{Step 3: calibrate at the meeting point.} The gathered branch input meets the logic embedding before, not after, the nonlinear transformation:
\begin{equation}
\mathbf{h}_0 = \mathbf{H}^{T} \mathbf{A}_m , \qquad
\mathbf{O} = \mathcal{T}(\mathbf{h}_0 + \mathbf{G}),
\end{equation}
where $\mathcal{T}$ is a linear layer. The order matters. Because $\mathbf{G}$ enters before $\mathcal{T}$, the transformation is computed on a representation that already carries the rule's decision; the network does not first form its own view and then get corrected, it forms its view around the anchor from the start.

\textbf{Step 4: return and settle.} The calibrated branch is routed back into the streams and the streams are pooled:
\begin{equation}
\hat{\mathbf{H}} = \mathbf{B}^{T}\mathbf{O}^{T} + \mathbf{A}_r^{T}\mathbf{H},
\qquad
\mathbf{h}_{\mathrm{final}} = \sum_{i=1}^{n} \hat{\mathbf{H}}_{i,:}.
\end{equation}
Each stream receives the calibrated signal in its own proportion $\beta_i$, keeps its laterally mixed residual, and the sum settles the streams into a single consolidated state. The policy and value heads read $\mathbf{h}_{\mathrm{final}}$, nothing else.

The full forward pass is summarized in Algorithm~\ref{alg:hcr}, and its integration with PPO in Algorithm~\ref{alg:ppo}: at every environment step, the state is embedded, the rule is consulted, HCR consolidates the two, and the resulting $\mathbf{h}_{\mathrm{final}}$ drives both the action and the value estimate; the PPO objectives are unchanged. The anchor is thus not a regularizer added to the loss, nor a supervisor added to the output. It is a structural fact of the forward pass, present in every decision the policy ever makes.

\section{Experiments}

\textbf{Setup.} All experiments run on an AMD Threadripper 7960X CPU and a single RTX A6000 (48\,GB) GPU. The agent selects from six bitrates $\{300, 750, 1200, 1850, 2850, 4300\}$\,kbps; chunks last
4\,s, the buffer caps at 60\,s, and each video has 49 chunks. For a fair comparison we inherit the full hyperparameter configuration of PA-MoE \cite{he2025plasticity} and change only the network architecture: PPO with learning rate $10^{-4}$, batch 2000, minibatch 62, 5 epochs, $\gamma=0.99$, $\lambda=0.95$, clip $\epsilon=0.2$, entropy 0, value coefficient 5, ReLU activations. Any performance difference is therefore attributable to the architecture alone.

\textbf{Baselines.} Three rule-based and three learning-based methods: \textbf{BufferBased} \cite{huang2014buffer} selects bitrates from buffer occupancy; \textbf{RateBased} \cite{li2014probe} picks the highest rate a throughput estimate sustains; \textbf{RobustMPC} \cite{yin2015control} optimizes QoE over a short horizon while hedging against prediction error; \textbf{Pensieve} \cite{mao2017neural} learns a DRL policy end-to-end; \textbf{Merina} \cite{kan2025merina+} meta-learns for fast adaptation; \textbf{PA-MoE} \cite{he2025plasticity} adapts through plasticity-aware mixture-of-experts.

\textbf{Traces.} Training uses \textbf{Train} \cite{he2025plasticity} only. Evaluation spans \textbf{Test} \cite{he2025plasticity}, \textbf{FCC-18} \cite{kan2025merina+}, \textbf{HSDPA} \cite{riiser2013commute}, \textbf{Oboe} \cite{huang2019comyco} (\textbf{Train}, \textbf{Test}, \textbf{FCC-18}, \textbf{HSDPA} and \textbf{Oboe} are  3G network), and, for the hardest setting, \textbf{Lumos4G}\cite{narayanan2021variegated}, \textbf{Lumos5G} \cite{narayanan2021variegated} and \textbf{SolisWi-Fi} \cite{GreenLv}, whose bandwidth scales differ from training by up to two orders of magnitude. We additionally construct \textbf{train\_uncorr} by reshuffling the \textbf{Train} sequences, preserving every marginal statistic while destroying temporal structure; it serves as the controlled texture-OOD probe throughout. Following Protocol~\ref{proto:tage}, we evaluate each checkpoint saved during training on all target datasets, running 300 independent episodes per checkpoint. This yields full generalization trajectories rather than single-point scores; bars average the last 15 evaluation steps, and CDFs aggregate all session-level outcomes \cite{he2025plasticity}.

\subsection{What Matters: the Rule or the Architecture?}
\label{TheArchitecturalBottleneckImpactofNeuralStructures}

\begin{figure}[!htbp]
\centering
\includegraphics[width=\columnwidth]{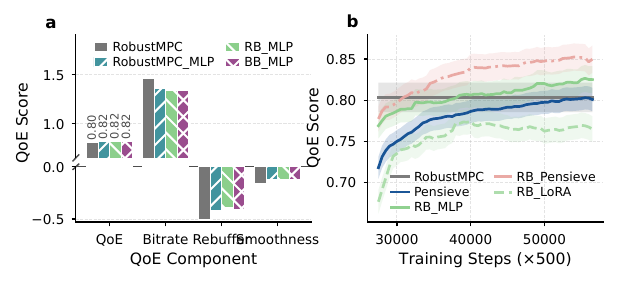}%
\caption{(a) QoE Performance: Comparing Different Rule-Based Approaches under MLP Integration. (b) Total QoE (3G) learning curves demonstrating the critical impact of different neural architectures under a fixed RB rule integration.}
\label{different_rule_based_mlp}
\end{figure}

Before comparing against the field, we ask which half of a neuro-symbolic design carries the weight. Two controlled sweeps answer it. First we fix the architecture to a plain MLP and vary the rule: RateBased (RB), BufferBased (BB), and RobustMPC priors yield nearly identical QoE ($0.80$--$0.82$; Fig.~\ref{different_rule_based_mlp}(a)). The choice of rule barely matters. Then we fix the rule to RateBased and vary the architecture: the outcome now swings from strong gains (RB\_Pensieve, above $0.85$) to outright negative transfer (RB\_LoRA; Fig.~\ref{different_rule_based_mlp} (b)). Where the choice of rule barely moved the result, the choice of architecture swings it from gain to harm. The value of a logic prior is thus real but conditional, and the condition is an architecture able to absorb it. HCR is our answer to that condition.

\subsection{Main Comparison}

\begin{figure}[!htbp]
\centering
\includegraphics[width=2.0in]{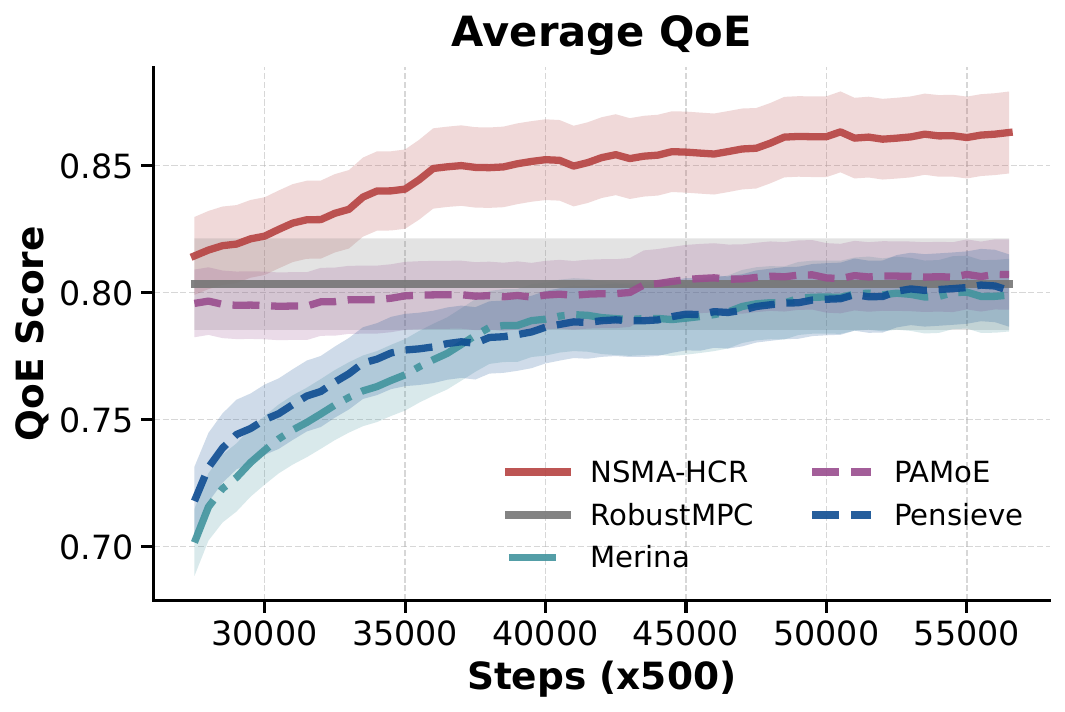}%
\caption{QoE learning curves comparing the proposed NSMA HCR framework against state-of-the-art baselines.}
\label{performance_total_qoe}
\end{figure}

Having built the architecture that the logic prior demands, we now ask the central question of this paper: trained on a single corpus, can one policy hold the rule's robustness and the network's adaptability? We compare NSMA against all six baselines under Protocol~\ref{proto:tage}. We begin on the five 3G-era traces that share the bandwidth scale of the training set (Train, Test, FCC-18, HSDPA, and Oboe), the common testbed of prior ABR work \cite{mao2017neural, kan2025merina+, he2025plasticity}, whose temporal properties are already characterized in Fig.~\ref{trace_question}. The harder evaluation settings, 4G, 5G, Wi-Fi, and real hardware, are examined in detail in a later section. Fig.~\ref{performance_total_qoe} tracks Total QoE on these traces against all baselines under Protocol~\ref{proto:tage}. RobustMPC sets the level that learning-based methods have struggled to pass. Pensieve, Merina, and PA-MoE all converge just below it (0.782--0.801 vs.\ 0.803; Table~\ref{tab:result}). NSMA-HCR leads throughout training and clears it, converging above 0.85.

Two further observations matter more than the final number. Together they show that the design keeps both of its promises: the policy starts strong because of the rules, and stays safe while it learns. First, there is no cold start. Pensieve and Merina open below 0.72 and must learn everything from scratch, while NSMA-HCR opens above 0.80, because the anchor carries what the rules have always known into the very first decision. Second, the lead costs no safety. Table~\ref{tab:result} shows NSMA-HCR reaches the bitrate of aggressive RL policies while keeping rebuffering and smoothness at rule-based levels. Normally, a higher bitrate is paid for with more stalls; here, that price is not paid. Finally, the two variants let us split the credit. NSMA-Add, which does nothing but place $\mathbf{G}$ inside the representation, already reaches 0.831 and beats every baseline, so the placement alone accounts for most of the gain. The routed calibration of NSMA-HCR then lifts it further to 0.848.

\begin{table}[!htbp]
  \centering
  \caption{QoE and its components for different ABR algorithms
           (mean over datasets and episodes).}
  \label{tab:result}
  \begin{tabular}{@{}l S S S S@{}}
    \toprule
    Algorithm & {QoE $\uparrow$} & {Bitrate $\uparrow$}
              & {Rebuffer $\downarrow$} & {Smoothness $\downarrow$} \\
    \midrule
    \multicolumn{5}{@{}l}{\textit{Rule-based}} \\
    \quad Buffer Based & 0.567 & 1.315 & -0.084 & -0.388 \\
    \quad Rate Based   & 0.519 & 1.490 & -0.147 & -0.339 \\
    \quad Robust MPC   & 0.803 & 1.462 & -0.116 & -0.159 \\
    \addlinespace
    \multicolumn{5}{@{}l}{\textit{Learning-based}} \\
    \quad Pensieve & 0.782 & 1.271 & -0.093 & -0.087 \\
    \quad MERINA   & 0.799 & 1.278 & -0.095 & -0.092 \\
    \quad PA-MoE   & 0.801 & 1.259 & -0.085 & -0.094 \\
    \addlinespace
    \multicolumn{5}{@{}l}{\textit{NSMA (Ours)}} \\
    \quad NSMA Add & 0.831          & 1.344 & -0.092 & -0.119 \\
    \quad NSMA HCR & \bfseries 0.848 & 1.337 & -0.090 & -0.103 \\
    \bottomrule
  \end{tabular}
\end{table}

\begin{figure*}[!htbp]
\centering
\includegraphics[width=1.35in]{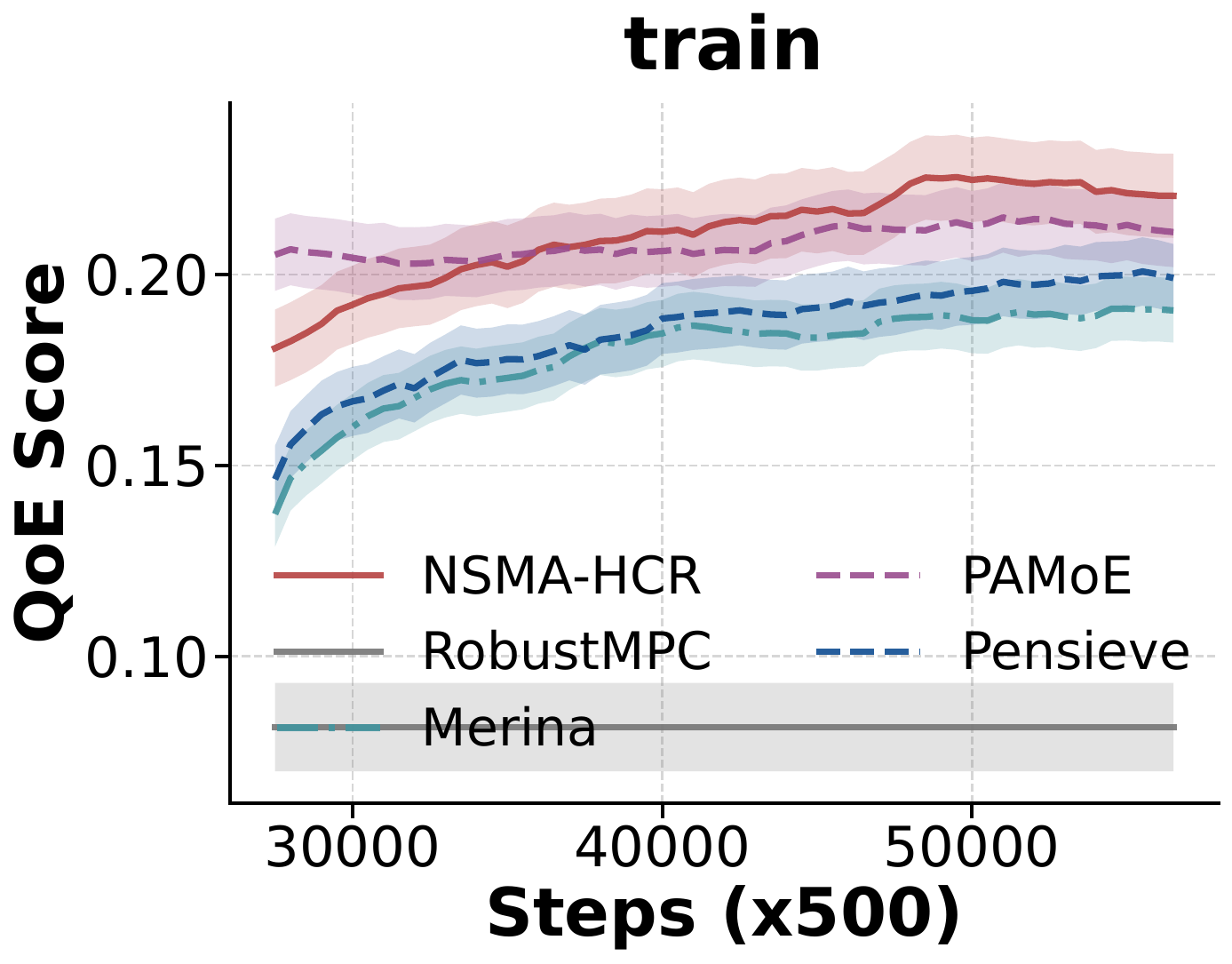}
\includegraphics[width=1.35in]{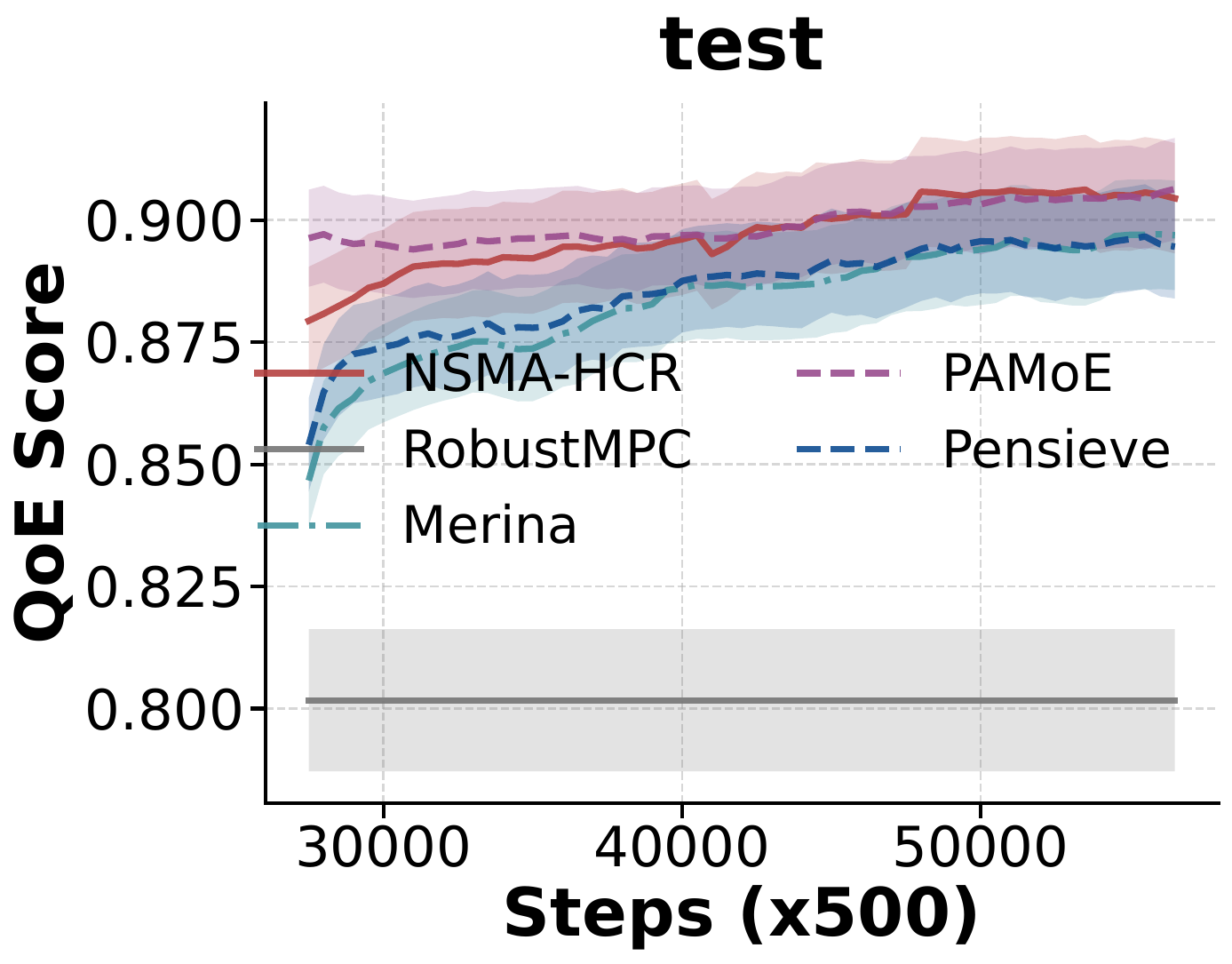}
\includegraphics[width=1.35in]{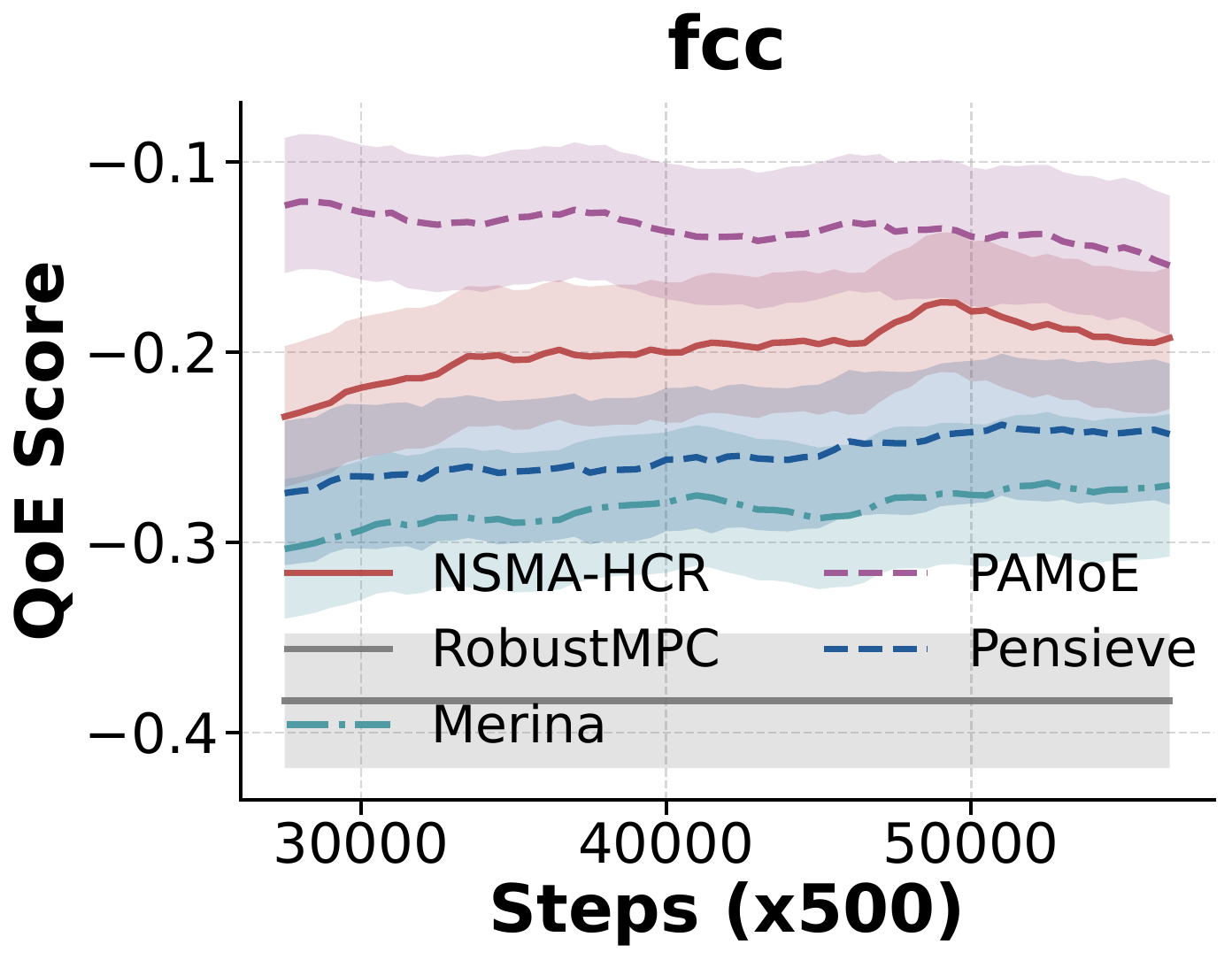}
\includegraphics[width=1.35in]{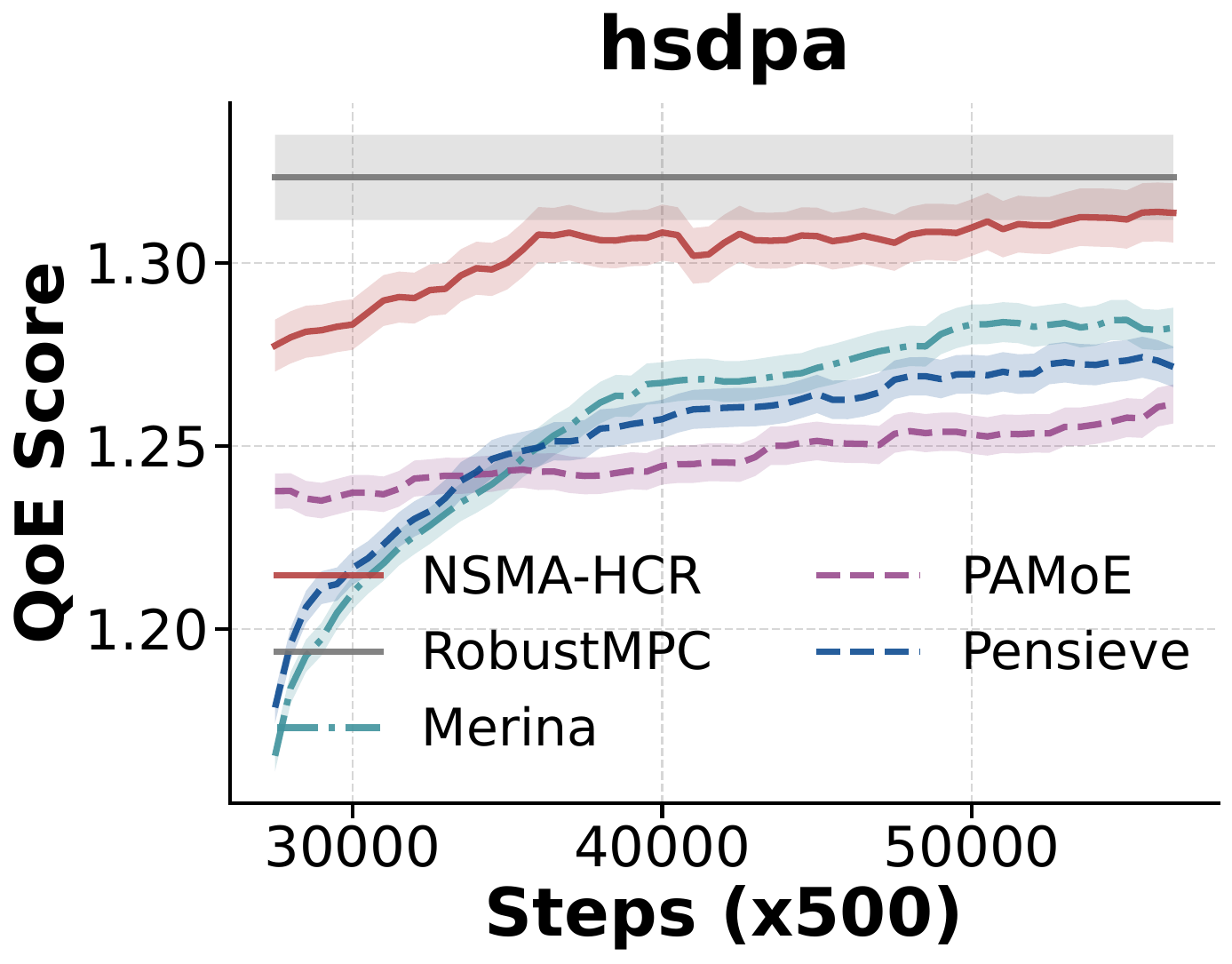}
\includegraphics[width=1.35in]{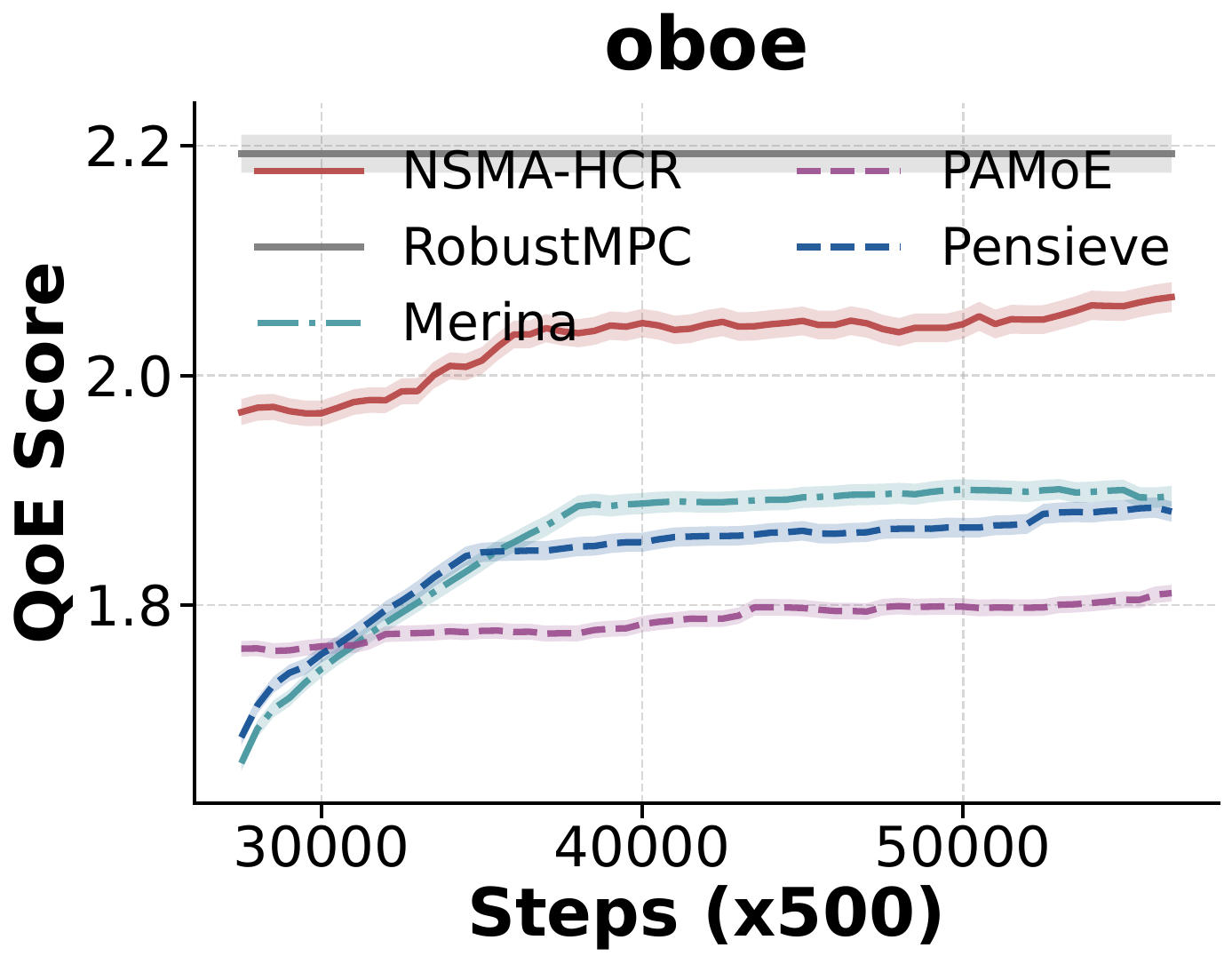}
\includegraphics[width=1.35in]{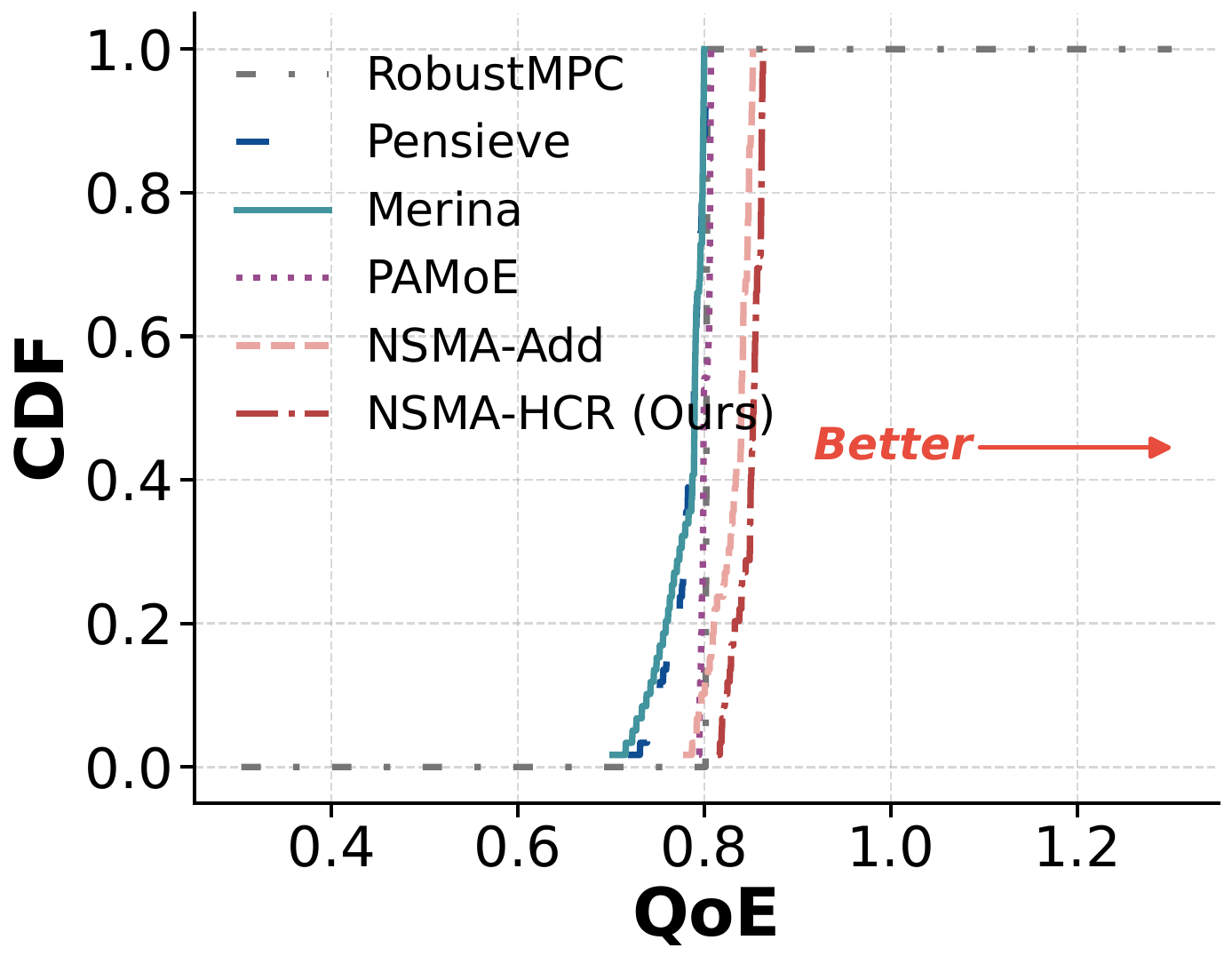}
\includegraphics[width=1.35in]{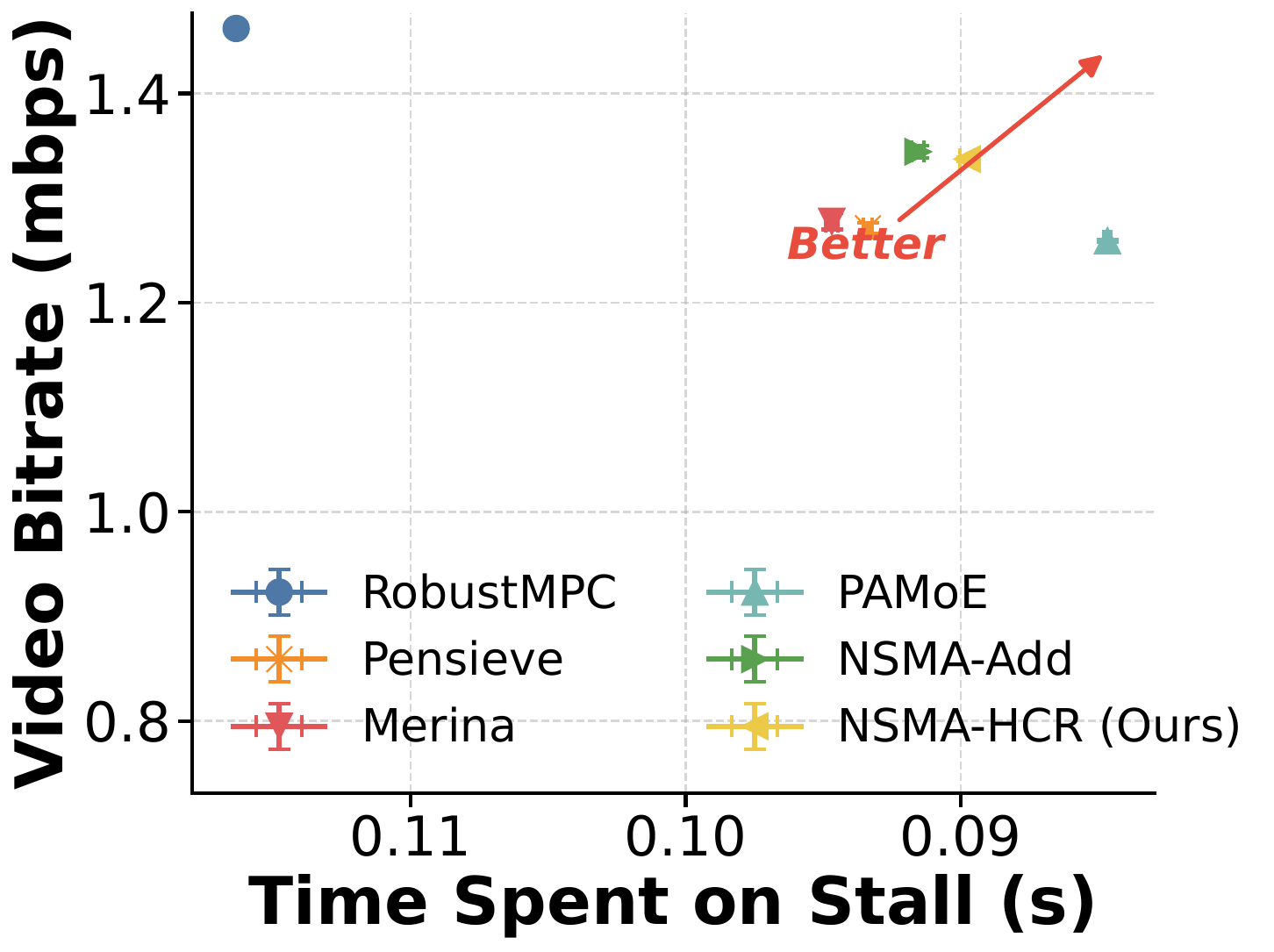}
\includegraphics[width=1.35in]{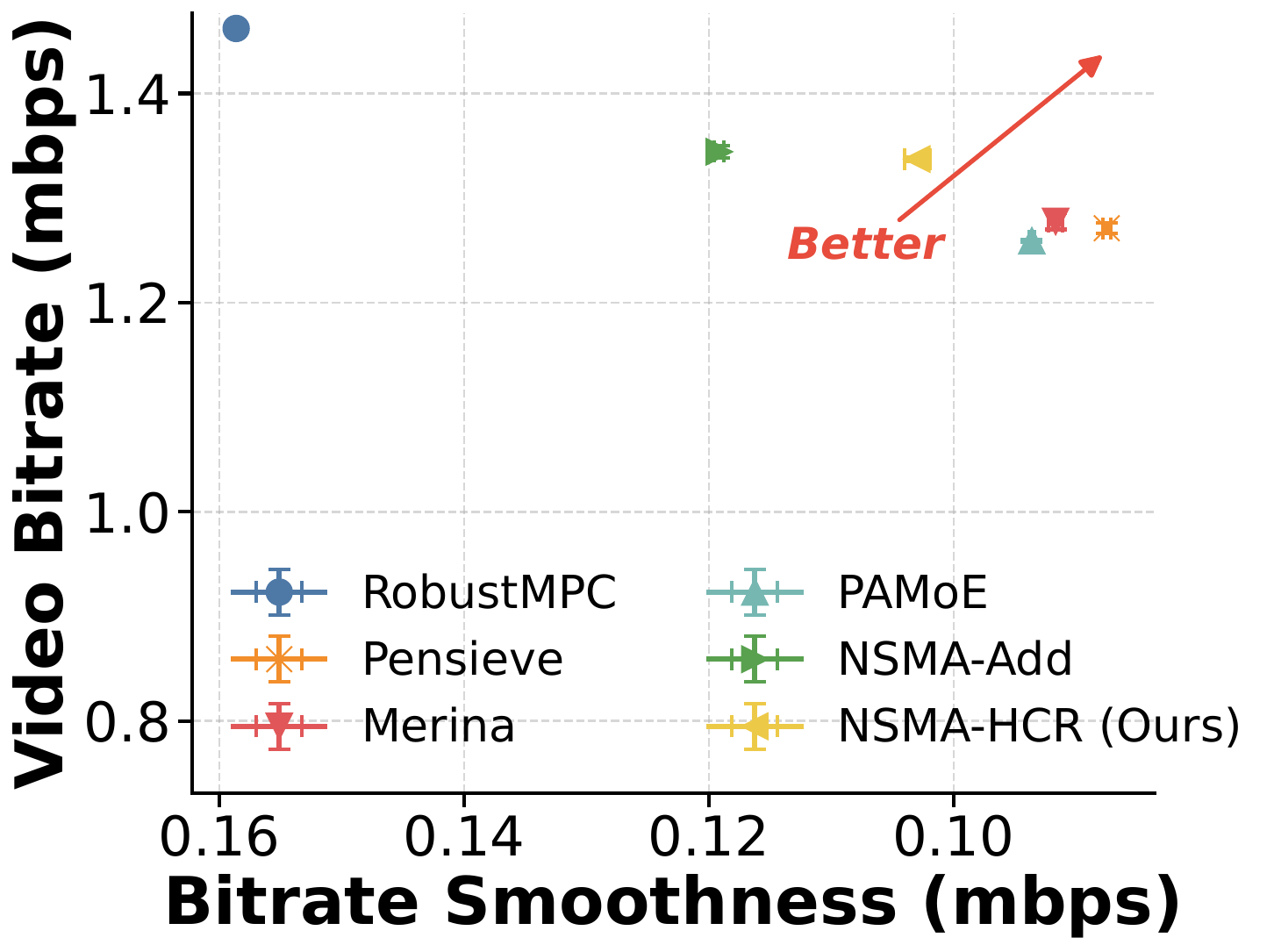}
\includegraphics[width=1.35in]{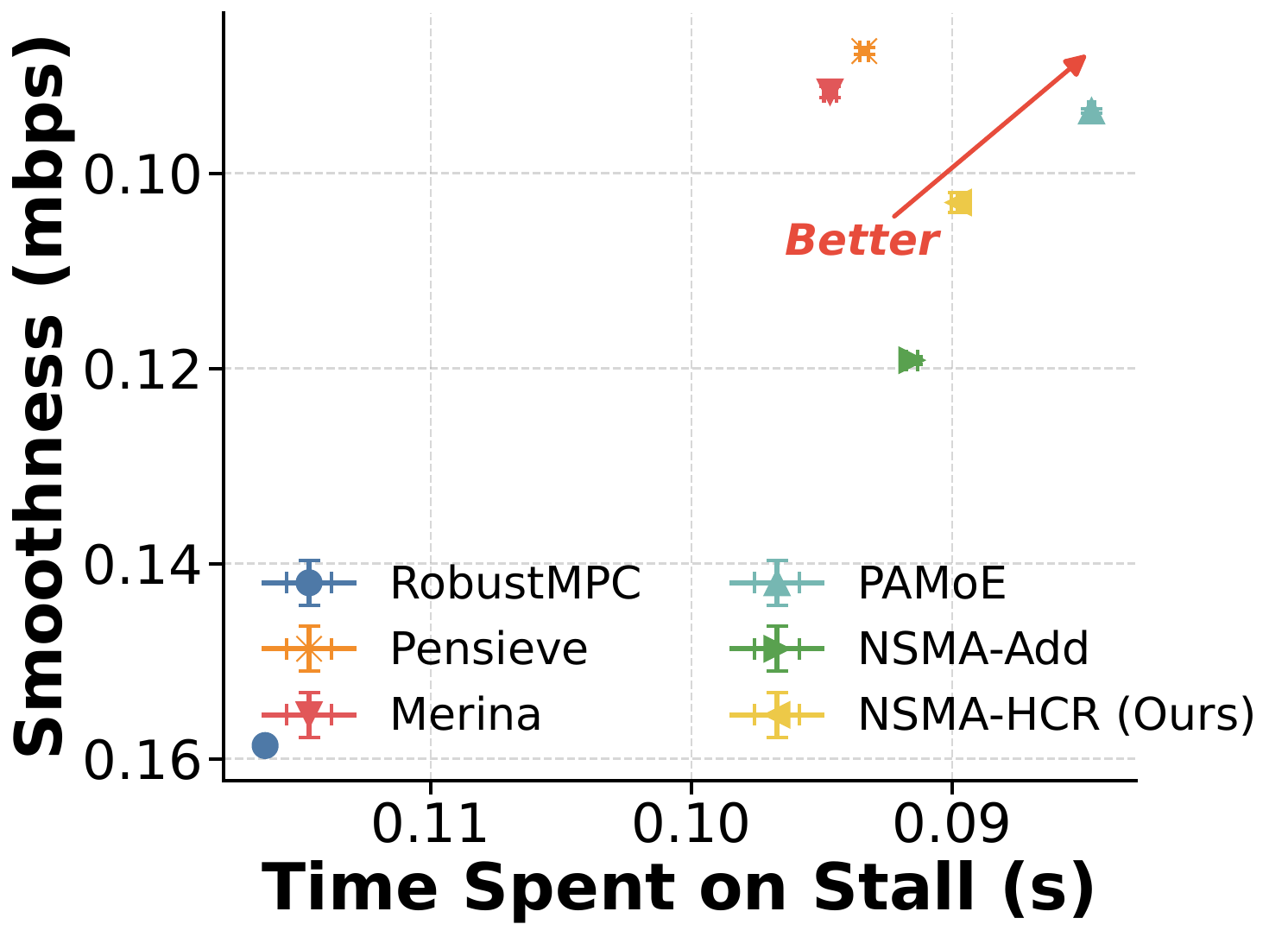}
\includegraphics[width=1.35in]{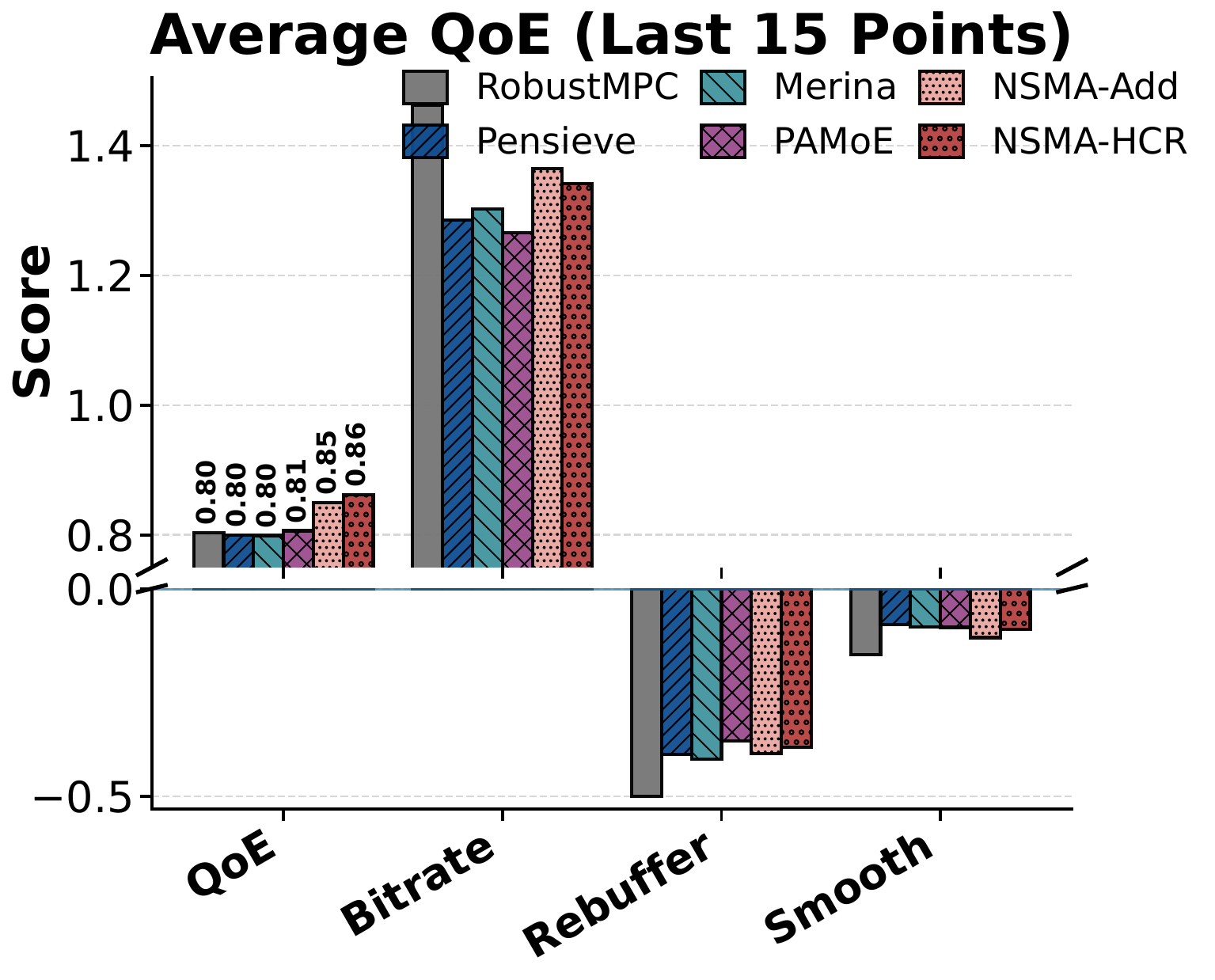}
\caption{NSMA leads across every evaluation trace and every view of QoE. \emph{Top:} per-dataset QoE trajectories under Protocol~\ref{proto:tage}; \emph{Bottom:} the aggregated CDF of total QoE, where NSMA-HCR reaches high QoE; pairwise trade-offs; and the QoE component breakdown, confirming the lead comes from bitrate held high without added stalls.}
\label{qoe_under_diff_algo_diverse_analysis}
\end{figure*}

The per-dataset trajectories in Fig.~\ref{qoe_under_diff_algo_diverse_analysis} (top) show where the lead comes from. It widens most on \texttt{hsdpa} and \texttt{oboe}, the two traces whose temporal structure departs furthest from training (Fig.~\ref{trace_question}). These are exactly the environments where previously learned policies fall behind the rules. The aggregated CDF (bottom left) shows that NSMA-HCR reaches high QoE in a larger share of sessions than any baseline, with NSMA-Add second. The trade-off panels tell the same story pairwise: for bitrate against stalls, and bitrate against smoothness, both variants sit where quality is high and penalties are low.

\subsection{Ablation: Which Part Does What}

The full model rests on two ingredients. One is the logic prior $\mathbf{G}$, the knowledge itself. The other is the HCR routing that carries this knowledge into the representation. To learn what each contributes, we remove one at a time and retrain. NSMA HCR w/o P keeps the routing but drops the prior; NSMA HCR w/o H keeps the prior but injects it by plain addition, with no routing at all.

\begin{figure}[!htbp]
\centering
\includegraphics[width=2.0in]{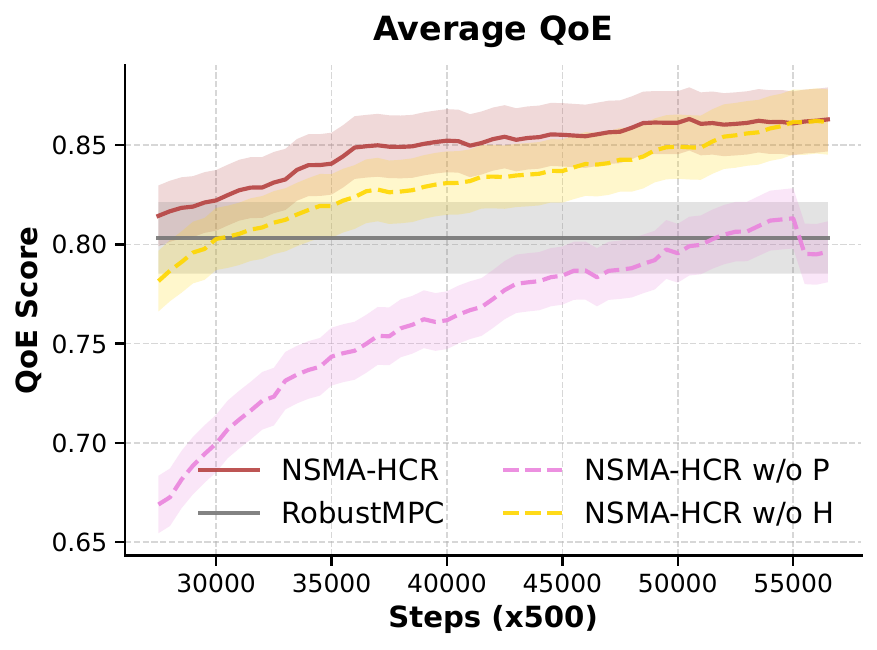}%
\caption{Ablation of NSMA-HCR.}
\label{ablation_w_o_l_performance}
\end{figure}

Fig.~\ref{ablation_w_o_l_performance} shows that each removal leaves a different kind of damage. Take away the prior, and the strong start disappears. The model opens from a low QoE like an ordinary RL agent. So the prior is what the model starts with, and also keeps it stable in the later stage of training. Take away the routing, and the knowledge is still added in, but the model cannot make good use of it at first. It still opens low, and only slowly climbs to where the full model stands. So the routing is what turns the prior from something the model has into something the model uses. The two kinds of damage match the two roles exactly. The prior provides the knowledge, the routing lets the model use it, and the full model needs both.

\subsection{Inside the Latent Space}
\label{sec:manifold}

\begin{figure}[!htbp]
\centering
\includegraphics[width=2.5in]{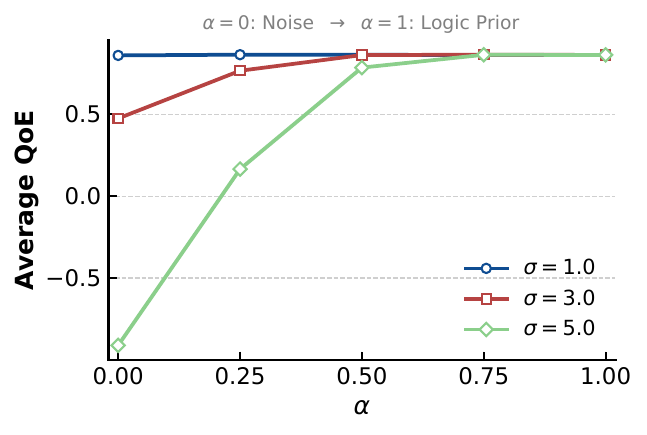}%
\caption{Perturbation test of the logic embedding. }
\label{manifold_analysis}
\end{figure}

Does HCR truly anchor the representation, or does it merely add a useful input? If the anchor is real, three things should follow. Small damage to $\mathbf{G}$ should not hurt much. Moving a damaged $\mathbf{G}$ back toward the real one should steadily restore performance. And the restoration should be gradual, with no sudden jumps. We test all three at once. We corrupt $\mathbf{G}$ with white noise at three strengths, $\sigma \in \{1.0, 3.0, 5.0\}$, and blend the corrupted and the real embedding as $\mathbf{G}_{\alpha} = \alpha \mathbf{G} + (1-\alpha)\boldsymbol{\epsilon}_{\sigma}$, sweeping $\alpha$ from pure noise to pure logic. Fig.~\ref{manifold_analysis} confirms all three predictions. When the noise is mild, replacing $\mathbf{G}$ entirely with the noisy version barely lowers QoE. 

At every noise strength, including the heaviest, QoE rises steadily as $\alpha$ grows. And every recovery curve is smooth from end to end. These curves, however, allow a simpler reading. Any useful input would produce the same shape, since more of a useful signal naturally means better performance. Under that reading, $\mathbf{G}$ helps only because it carries information, and the anchor plays no part. What can rule it out is an experiment that keeps the information identical but breaks the placement, and we run exactly that experiment in Section \ref{sec:realworld}.

\subsection{The Strictest Test: Foreign Networks and Real Hardware}
\label{sec:realworld}

\textbf{Foreign networks.} Everything so far ran on traces that share the bandwidth scale of training. We now remove that comfort. The trained policy is deployed, with no fine-tuning, on Lumos4G (175 traces, mean 40.4\,Mbps), Lumos5G (121 traces, 522.9\,Mbps), and SolisWi-Fi (80 traces, 24.2\,Mbps), networks whose bandwidth exceeds training by up to two orders of magnitude. The three also split cleanly along the axis this paper cares about. Lumos4G and 5G are foreign in scale but familiar in texture, keeping strong short-range correlation ($0.885$ and $0.900$ at lag~1). SolisWi-Fi is foreign in both scale and texture, close to white noise ($0.197$ at lag~1, $0.062$ at lag~10), leaving almost nothing for a learned policy to predict.

\begin{figure}[!htbp]
  \centering
  \begin{subfigure}[b]{0.48\columnwidth}
    \centering
    \includegraphics[width=\linewidth]{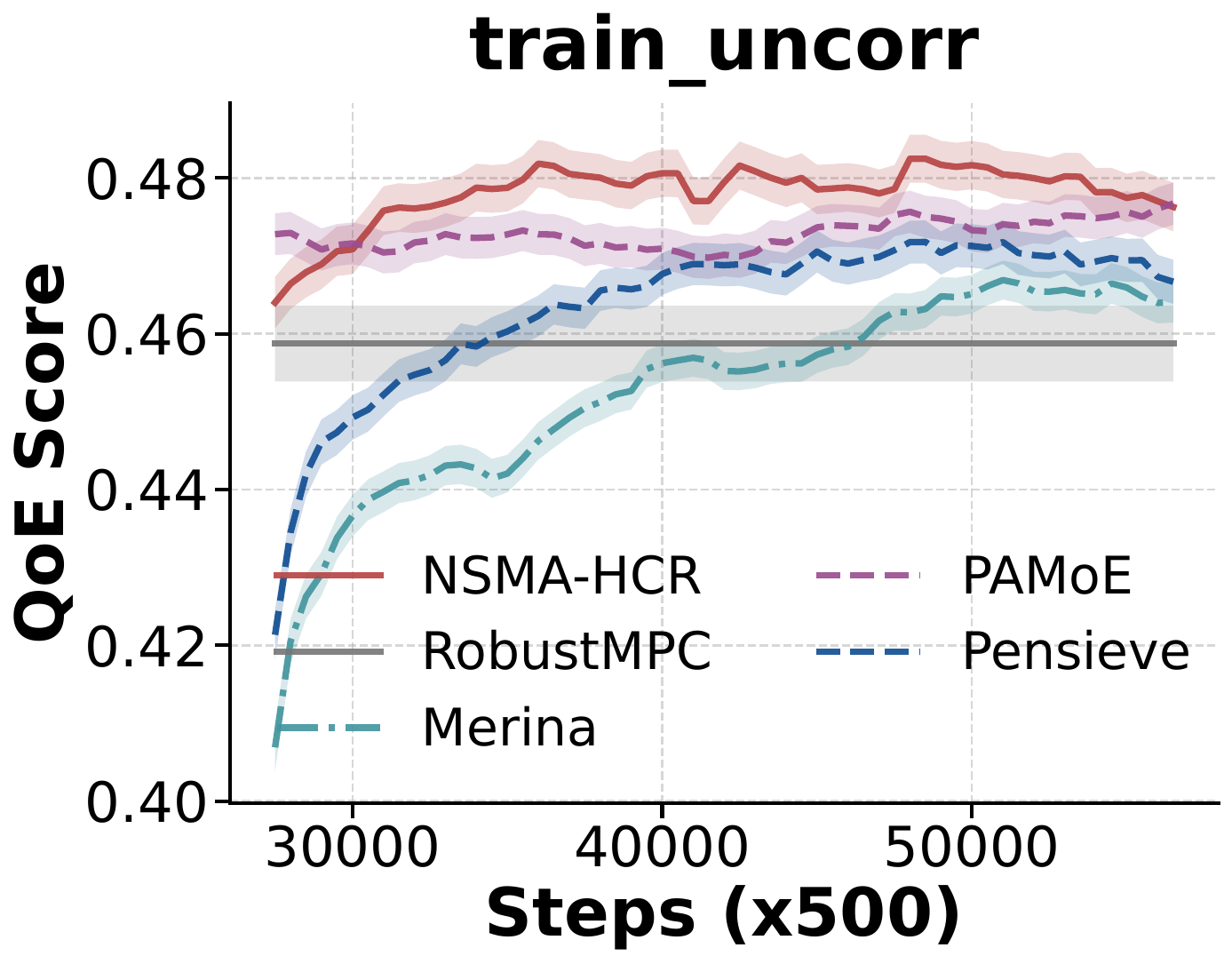}
    \label{fig:train_uncorr}
  \end{subfigure}\hfill
  \begin{subfigure}[b]{0.48\columnwidth}
    \centering
    \includegraphics[width=\linewidth]{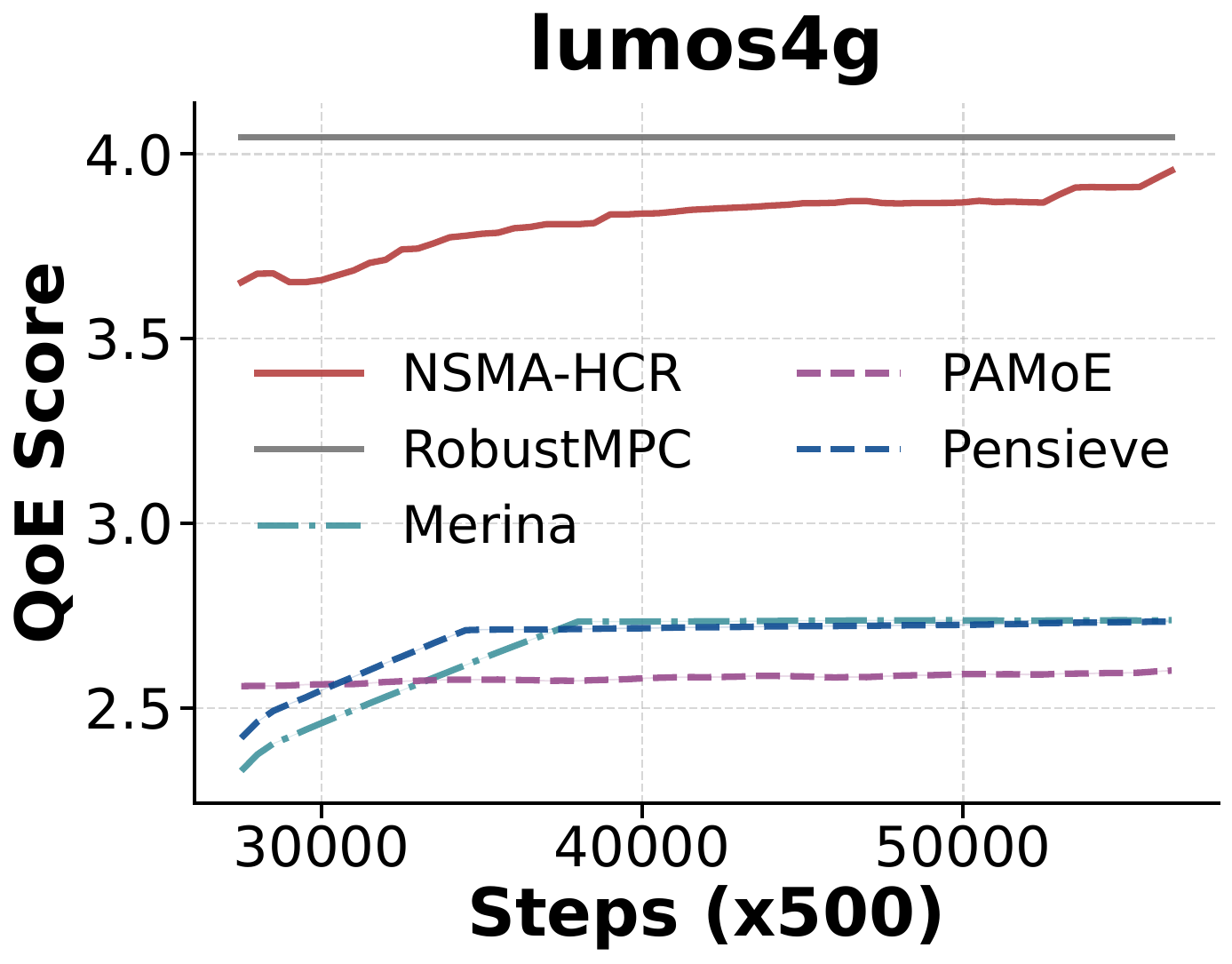}
    \label{fig:lumos4g}
  \end{subfigure}\\[0.5em]
  \begin{subfigure}[b]{0.48\columnwidth}
    \centering
    \includegraphics[width=\linewidth]{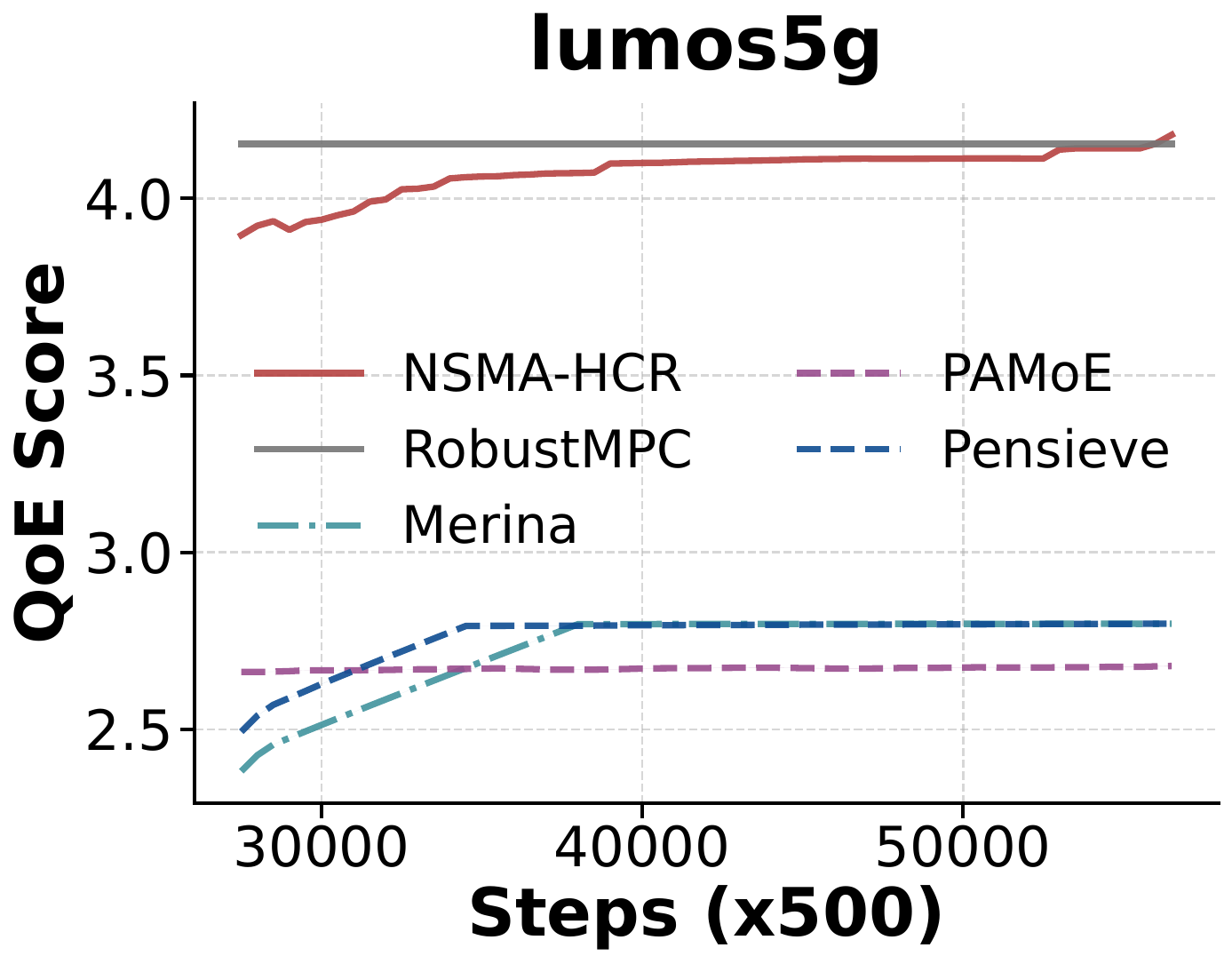}
    \label{fig:lumos5g}
  \end{subfigure}\hfill
  \begin{subfigure}[b]{0.48\columnwidth}
    \centering
    \includegraphics[width=\linewidth]{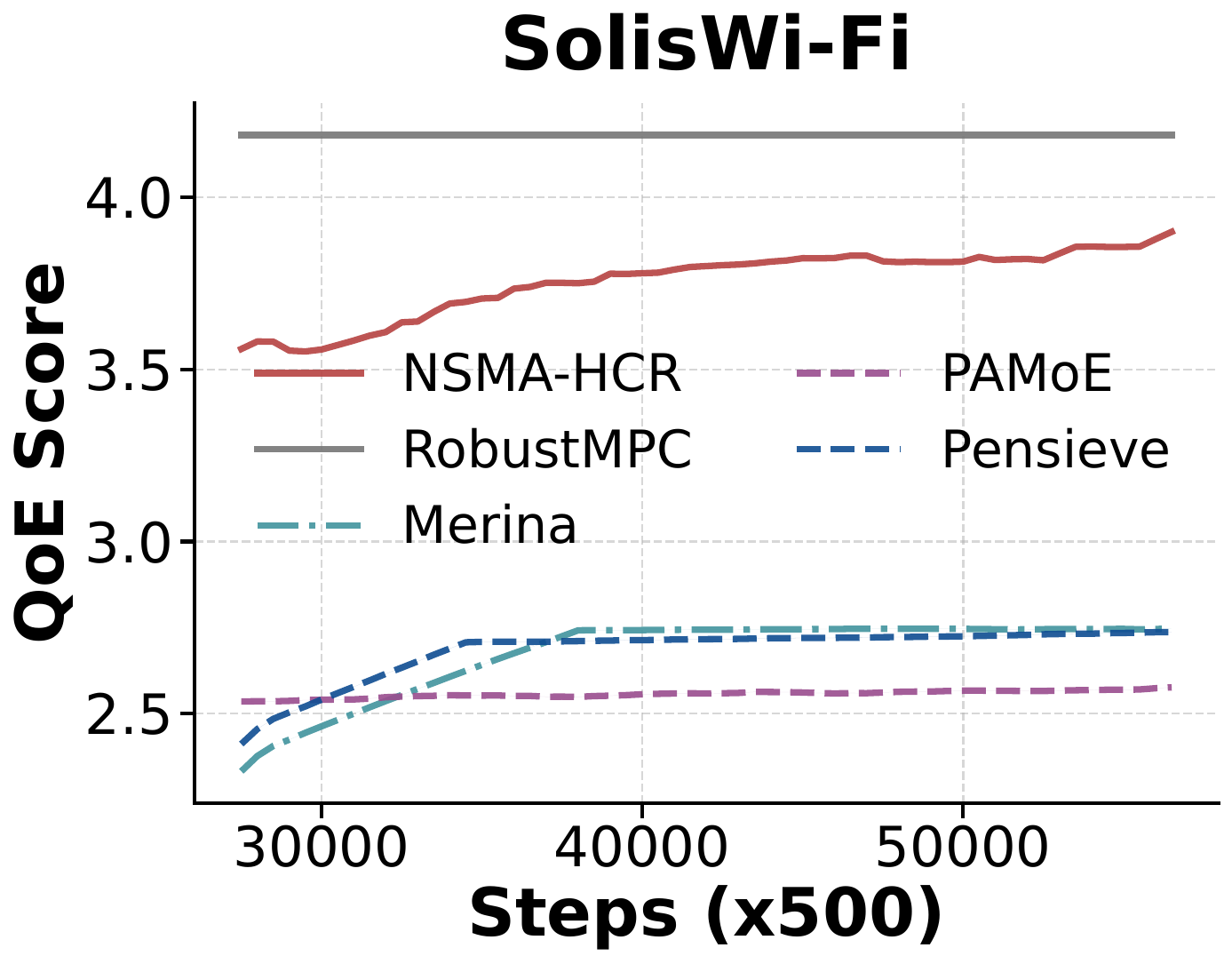}
    \label{fig:solis_wifi}
  \end{subfigure}
  \caption{QoE on heterogeneous networks, without fine-tuning.}
  \label{fig:performance_heterogeneous_network}
\end{figure}

Fig.~\ref{fig:performance_heterogeneous_network} shows the results on these three networks. The rule-based methods perform best here, as expected. The existing learned policies, Pensieve, Merina, and PA-MoE, fall far behind them. NSMA is the only learned policy that stays close to the rule-based level, and it does so on all three networks. 

Across all three foreign networks, the pattern is uniform. The existing learned policies flatten out around the same level, far below the rule-based methods, while NSMA-HCR climbs close to the rule-based line on every one of them. The \texttt{train\_uncorr} panel adds a further point. On this trace, Section~\ref{sec:audit} showed that the advantage of learned policies disappears entirely; NSMA-HCR is the one method that ends clearly above RobustMPC. What a policy learns from the training traces does not carry over to networks. What the rule prior provides does.

\begin{figure}[!htbp]
\centering
\includegraphics[width=\columnwidth]{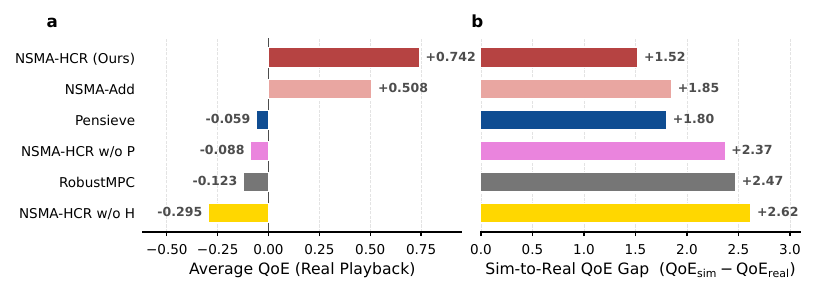}%
\caption{Real-world testbed validation.}
\label{realworld_summary}
\end{figure}

\textbf{Real hardware.} One doubt survives every simulator result, namely that the policy may have mastered the simulator rather than the problem. So we take the last step. Each method runs 27 (3 $\times$ 9 traces) sessions of real HLS playback over 3 seeds, on an unmodified \texttt{hls.js} player in headless Chromium, with bandwidth throttled in real time by the traces and QoE measured under the same model as before. Nothing is retrained, nothing is adjusted. Real playback proves harsher than any simulation. On real playback, every baseline ends with a negative QoE, even RobustMPC. The real player brings extra costs that the simulator does not have, such as request delays, buffering behavior, and rendering, and these costs pull every existing method below zero (Fig.~\ref{realworld_summary} (a)). Only the two NSMA variants stay positive.

Fig.~\ref{realworld_summary} (b) then answers the doubt we began with. For each method we compute the sim-to-real gap, the absolute difference between its simulator QoE and its real QoE on paired runs. Every method degrades when it leaves the simulator. NSMA-HCR degrades the least. Together the two panels close the argument from both sides. NSMA-HCR performs best on real hardware, and its advantage is not tied to the simulator, since it loses the least when moving from simulation to real playback.

\textbf{Representation analysis on real hardware.} The latent-space analysis in Section~\ref{sec:manifold} left one question open: whether $\mathbf{G}$ helps through its placement in the representation or simply as an informative input. We address it here on states collected from real playback. States from real sessions on \texttt{train} and \texttt{train\_uncorr} are passed through the frozen actor three times, each time with a different version of $\mathbf{G}$.

\begin{figure}[!htbp]
\centering
\includegraphics[width=\columnwidth]{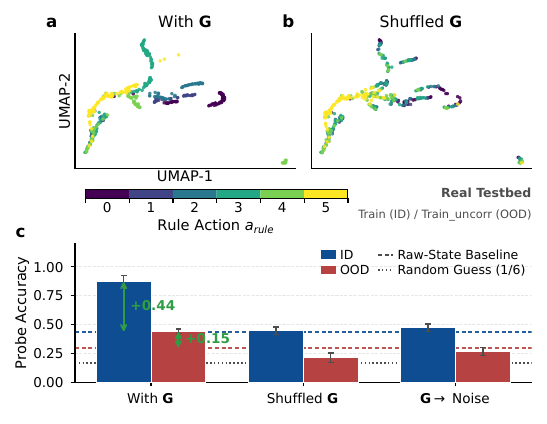}%
\caption{The rule action space guides the actor's representation manifold.}
\label{manifold_umap_real}
\end{figure}

With the true $\mathbf{G}$, a UMAP \cite{mcinnes2018umap} projection of the actor's internal features, colored by rule action, is cleanly ordered along the bitrate ladder on both datasets (Fig.~\ref{manifold_umap_real} (a)). We then shuffle $\mathbf{G}$ across states. This keeps the set of injected vectors, and hence the information content, unchanged; it only breaks the match between each state and its rule action. The ordering within each branch collapses (Fig.~\ref{manifold_umap_real} (b)). If $\mathbf{G}$ helped only as information, shuffling would have no effect.

To measure this effect with a number, we use a simple test. We train nothing; we only ask, for each state, whether its nearest neighbors in the actor's internal feature space share the same rule action, using a 10-nearest-neighbor classifier \cite{belinkov2022probing}. If states with the same rule action sit close together inside the representation, this classifier scores high; if they are scattered, it scores low. The score therefore tells us how strongly the rule is written into the representation. One care is needed in choosing the reference. The rule action is computed from the state, so even without $\mathbf{G}$, the raw state already predicts it to some degree. The fair reference is thus what the raw state alone achieves, not random guessing. Against this reference, adding the true $\mathbf{G}$ raises the score from $0.433$ to $0.872$ on \texttt{train}. In other words, without $\mathbf{G}$, fewer than half of a state's nearest neighbors are assigned the same rule action as the state itself. With $\mathbf{G}$, almost all of them are. On \texttt{train\_uncorr}, the trace where the learned advantage was shown to vanish, this clustering by rule action is weaker than on \texttt{train}, but it is still clearly present and well above the raw-state level (Fig.~\ref{manifold_umap_real} (c)).

\section{Conclusion and Future Work}

This paper began with a contest that ABR has lived with for decades, between rules that never learn and networks that never stop forgetting. We showed first that the field has been measuring this contest with the wrong instrument. Dataset identity does not predict generalization difficulty, and we traced the true difficulty to Trace Texture, the temporal structure that statistics leave unrecorded. We codified the repair as a texture-aware evaluation protocol, and followed it ourselves. We then dissolved the contest rather than settling it. NSMA embeds the rule's decision inside the latent space of the neural policy, where perturbation, interpolation, and real-hardware analyses all find it doing what it was placed there to do, holding the representation steady across texture shifts that break every learned baseline. Trained on a single 3G corpus, the resulting policy generalizes to 4G, 5G, Wi-Fi, and an unmodified real player without fine-tuning.

This work has three limitations, and each points to a concrete next step. First, Trace Texture is a concept rather than a measurable quantity; we can show its effects, but we cannot yet compute it for a given trace. Developing such a measure would let the evaluation protocol quantify, not just report, the difficulty of a benchmark. Second, our anchor comes from a single family of rules, and whether other forms of expert knowledge serve as well is untested. Congestion control, scheduling, and caching all hold decades of distilled principles waiting for the same treatment. Third, three separate analyses support our geometric explanation of the anchor's benefit, but they show that the geometry and the benefit appear together, not that one causes the other. A formal account of when in-representation placement succeeds would turn this observation into a guarantee. None of these limits, however, is specific to video. Wherever hand-crafted principles and learned policy still stand on opposite sides of a boundary, the same placement is waiting to be tried.

\bibliographystyle{IEEEtran}
\bibliography{reference}

\end{document}